\documentclass[conference]{IEEEtran}
\usepackage{epsfig,endnotes}
\usepackage{amsmath}
\usepackage{graphicx}

\PassOptionsToPackage{hyphens}{url}\usepackage{hyperref}
\expandafter\def\expandafter\UrlBreaks\expandafter{\UrlBreaks%  save the current one
  \do\a\do\b\do\c\do\d\do\e\do\f\do\g\do\h\do\i\do\j%
  \do\k\do\l\do\m\do\n\do\o\do\p\do\q\do\r\do\s\do\t%
  \do\u\do\v\do\w\do\x\do\y\do\z\do\A\do\B\do\C\do\D%
  \do\E\do\F\do\G\do\H\do\I\do\J\do\K\do\L\do\M\do\N%
  \do\O\do\P\do\Q\do\R\do\S\do\T\do\U\do\V\do\W\do\X%
  \do\Y\do\Z}

\usepackage{tabularx}
\usepackage{acro}
\usepackage{listings}
\usepackage{tikz}
\usetikzlibrary{snakes}
\usepackage{subcaption}
\usepackage{multirow}
\usepackage{booktabs}
\usepackage[font={small,it}]{caption}
\usepackage{booktabs}
\usepackage{soul}

\usepackage{color, colortbl}
\definecolor{Gray}{gray}{0.9}
\definecolor{LightCyan}{rgb}{0.2,0.2,0.2}

\lstnewenvironment{teX}[1][]
      {\lstset{language=[LaTeX]TeX}\lstset{escapeinside={(*@}{@*)},
       numbers=left,numberstyle=\normalsize,stepnumber=1,numbersep=5pt,
       breaklines=true,
           framesep=5pt,
           basicstyle=\normalsize\ttfamily,
           showstringspaces=false,
           keywordstyle=\itshape\color{blue},
           stringstyle=\color{maroon},
        commentstyle=\color{black},
        rulecolor=\color{black},
        xleftmargin=0pt,
        xrightmargin=0pt,
        aboveskip=\medskipamount,
        belowskip=\medskipamount,
               backgroundcolor=\color{white}, #1
    }}
    {}

\newcounter{note}

\begin{document}
%\title{Tech Support Scam 2.0: Exposing Search-and-Ad Abuse Tactics and Infrastructure} % TODO: replace with your title
\title{By Hook or by Crook: Exposing the Diverse Abuse Tactics of Technical Support Scammers} % TODO: replace with your title
%\author{Bharat Srinivasan, Athanasios Kountouras, Najmeh Miramirkhani, Monjur Alam, Nick Nikiforakis, Manos Antonakakis, Mustaque Ahamad.}
\author{
\IEEEauthorblockN{Bharat Srinivasan}
\IEEEauthorblockA{School of Computer Science \\
Georgia Institute of Technology\\
bharat.srini@gatech.edu}\\
\IEEEauthorblockN{Monjur Alam}
\IEEEauthorblockA{School of Computer Science\\
Georgia Institute of Technology\\
malam31@gatech.edu}
\and

\IEEEauthorblockN{Athanasios Kountouras}
\IEEEauthorblockA{School of Computer Science \\
Georgia Institute of Technology\\
kountouras@gatech.edu}\\
\IEEEauthorblockN{Nick Nikiforakis}
\IEEEauthorblockA{Department of Computer Science\\
Stony Brook University\\
nick@cs.stonybrook.edu}\\
\IEEEauthorblockN{Mustaque Ahamad}
\IEEEauthorblockA{School of Computer Science \\
Georgia Institute of Technology\\
mustaq@cc.gatech.edu}

\and

\IEEEauthorblockN{Najmeh Miramirkhani}
\IEEEauthorblockA{Department of Computer Science\\
Stony Brook University\\
n.miramirkhani@stonybrook.edu}\\
\IEEEauthorblockN{Manos Antonakakis}
\IEEEauthorblockA{School of Electrical and Computer \\ Engineering\\
Georgia Institute of Technology\\
manos@gatech.edu}}

\maketitle

\begin{abstract}
Technical Support Scams (TSS), which combine online abuse with social engineering over the phone channel, have persisted despite several law enforcement actions. The tactics used by these scammers have evolved over time and they have targeted an ever increasing number of technology brands. Although recent research has provided important insights into TSS, these scams have now evolved to exploit ubiquitously used online services such as search and sponsored advertisements served in response to search queries. We use a data-driven approach to understand search-and-ad abuse by TSS to gain visibility into the online infrastructure that facilitates it. By carefully formulating tech support queries with multiple search engines, we collect data about both the support infrastructure and the websites to which TSS victims are directed when they search online for tech support resources. We augment this with a DNS-based amplification technique to further enhance visibility into this abuse infrastructure. By analyzing the collected data, we provide new insights into search-and-ad abuse by TSS and reinforce some of the findings of earlier research. Further, we demonstrate that tech support scammers are (1) successful in getting major as well as custom search engines to return links to websites controlled by them, and (2) they are able to get ad networks to serve malicious advertisements that lead to scam pages. Our study period of approximately eight months uncovered over 9,000 TSS domains, of both passive and aggressive types, with minimal overlap between sets that are reached via organic search results and sponsored ads. Also, we found over 2,400 support domains which aid the
TSS domains in manipulating organic search results. Moreover, to our surprise, we found very little overlap with domains that are reached via abuse of domain parking and URL-shortening services which was investigated previously. Thus, investigation of search-and-ad abuse provides new insights into TSS tactics and helps detect previously unknown abuse infrastructure that facilitates these scams.

\end{abstract}

% TODO: replace this section with code generated by the tool at https://dl.acm.org/ccs.cfm
%\begin{CCSXML}
%<ccs2012>
%<concept>
%<concept_id>10002978.10003029.10011703</concept_id>
%<concept_desc>Security and privacy~Usability in security and privacy</concept_desc>
%<concept_significance>500</concept_significance>
%</concept>
%</ccs2012>
%\end{CCSXML}

%\ccsdesc{Security and privacy~Use https://dl.acm.org/ccs.cfm to generate actual concepts section for your paper}
% -- end of section to replace with generated code

%\keywords{social engineering, fraud, tech support, cross-channel, telephony} % TODO: replace with your keywords

%\input{ccs-body} % TODO: replace with your brilliant paper!
\section{Introduction}
\label{section:introduction}

\begin{figure*}[h]
\centering
\resizebox{.9\linewidth}{!}{
\begin{tikzpicture}[snake=zigzag, line before snake = 5mm, line after snake = 5mm]
    % draw horizontal line  
    \draw[snake] (-1.5,0) -- (0,0); 
    \draw (0,0) -- (4.5,0);
    \draw[snake] (4.5,0) -- (6,0);
    \draw (6,0) -- (15.5,0);
    \draw[snake] (15.5,0) -- (17,0);

    % draw vertical lines
    \foreach \x in {0,3.5,7,9.5,13,15.5}
      \draw (\x cm,5pt) -- (\x cm,-10pt);

    % draw nodes
    \draw (0,0) node[below=10pt, text width=2cm] {\centering FTC cites use of SEO and ADs in TSS.~\cite{ftc_tss}} node[above=10pt] {\textbf{Jan 2014}};
    \draw (3.5,0) node[below=10pt, text width=2cm] {\centering Google and FB highlight difficulty in take downs of TSS due to the use of phone channel.~\cite{trustinads_report}} node[above=10pt] {\textbf{May 2014}};
    \draw (7,0) node[below=10pt, text width=2cm] {\centering 15 million Ads 25,000 websites blocked. TSS top list of bad actors in Search Engines~\cite{bing_bad_ads_report}} node[above=10pt] {\textbf{May 2, 2016}};
    \draw (9.5,0) node[below=10pt, text width=2cm] {\centering Bing Ads bans ads from third-party tech support services~\cite{bing_tsad_ban2}} node[above=10pt] {\textbf{May 12, 2016}};
    \draw (13,0) node[below=10pt, text width=2cm] {\centering FB Customer Service Scam on Google~\cite{fbsearch}} node[above=10pt] {\textbf{Jan 2017}};
    \draw (15.5,0) node[below=10pt, text width=2cm] {\centering Tech support scams persist with novel techniques~\cite{technetblog}} node[above=10pt] {\textbf{Apr 2017}};
    %\draw (3,0) node[below=3pt] {$  $} node[above=3pt] {$  $};
    %\draw (4,0) node[below=3pt] {$ 5 $} node[above=3pt] {$ 50 $};
    %\draw (5,0) node[below=3pt] {$ 6 $} node[above=3pt] {$ 60 $};
    %\draw (6,0) node[below=3pt] {$  $} node[above=3pt] {$  $};
    %\draw (7,0) node[below=3pt] {$ n $} node[above=3pt] {$ 10n $};
\end{tikzpicture}}
\caption{Timeline of some news events related to search-based technical support scams (TSS). \label{fig:news_timeline}}
\vspace{-3ex}
\end{figure*}
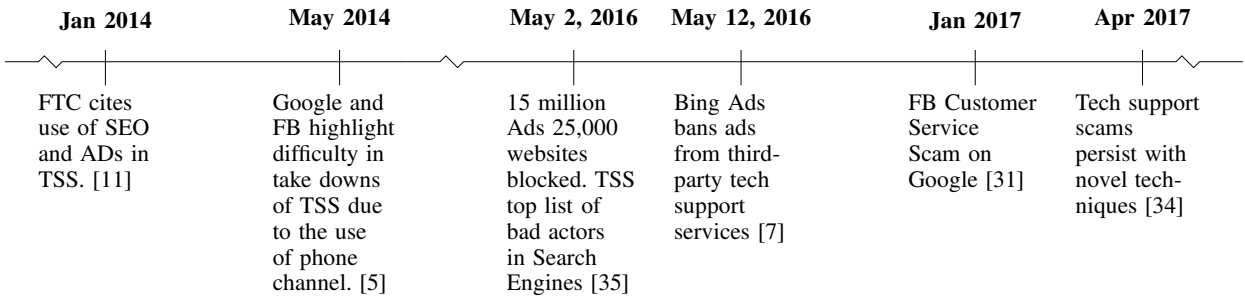

\begin{figure} [h]
    \centering
    \includegraphics[width=0.80\columnwidth,height=1.2\columnwidth]{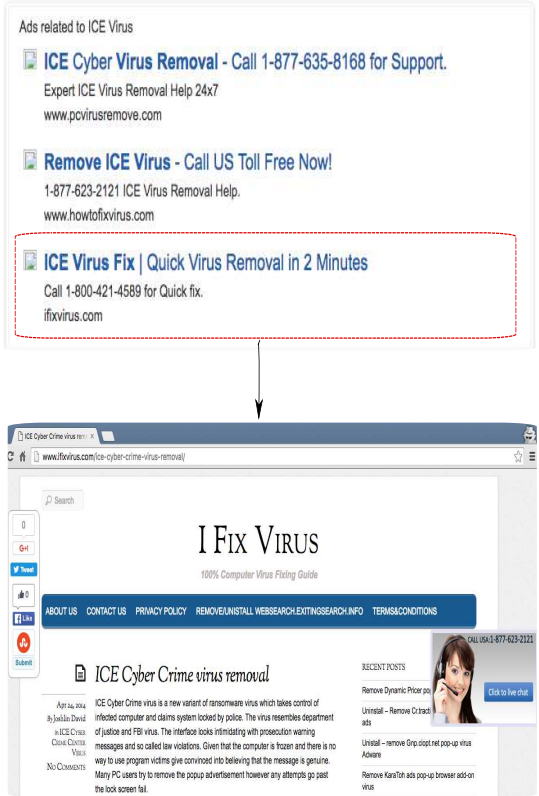}
    \caption{Screenshot from Goentry.com on July 1st, 2016.}
    \vspace{-4ex}
    \label{fig:goentryads}
\end{figure}

\begin{figure} [h]
    \centering
    \includegraphics[width=0.95\columnwidth,height=1.2\columnwidth]{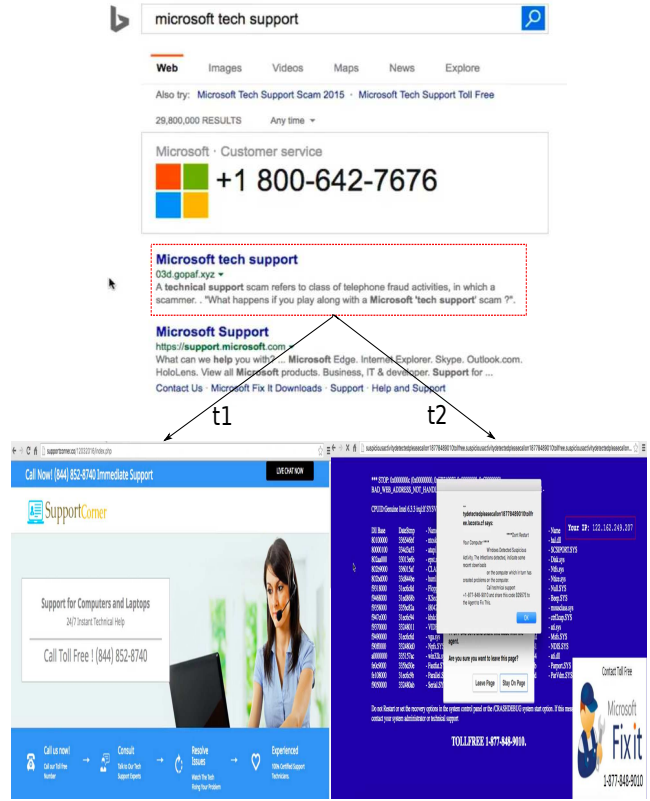}
    \caption{Screenshot from Bing.com on Feb 2nd, 2017.}
    \vspace{-4ex}
    \label{fig:bingads}
\end{figure}

The {\em Technical Support Scam} (TSS), in which scammers dupe their victims into sending hundreds of dollars for fake technical support services, is now almost a decade old. It started with scammers making cold calls to victims claiming to be a legitimate technology vendor but has now evolved into the use of sophisticated online abuse tactics to get customers to call phone numbers that are under the control of the scammers. 

In their pioneering research on TSS~\cite{miramirkhani2017dial}, Miramirkhani et. al. explored both the web infrastructure used by tech support scammers and the tactics used by them when a victim called a phone number advertised on a TSS website. They focused on TSS websites reached via malicious advertisements that are served by abusing domain parking and ad-based URL shortening services. Although their work provided important insights into how these services are abused by TSS, it has recently become clear that tech support scammers are diversifying their methods of reaching victims and the ways with which they convince these victims to call them on their advertised phone numbers.

Specifically, recent reports by the FTC and by search engines vendors suggest that scammers are turning to search engine results and the ads shown on search-results pages as novel ways of reaching victim users~\cite{ftc_tss,fbsearch,trustinads_report}. These new channels not only allow them to reach a wider audience but also allow them to diversify the ways with which they attempt to convince users to call them. As shown in Figure~\ref{fig:news_timeline}, several actions have been taken to stop TSS but these scams continue to adapt and evade both law enforcement and technical safeguards.

In this paper, we perform the first systematic study of these novel search-and-ad abuse channels. We develop a model for generating tech-support related queries and use the resulting 2,600 queries as daily searches in popular and less popular search engines. By crawling the organic search results and ads shown as a response to our queries (note that we follow a methodology that allows us to visit the websites of ads without participating in click-fraud), we discover thousands of domains and phone numbers associated with technical support scams. In addition to the traditional \emph{aggressive} variety of technical support scams (where the pages attempt to scare users into calling them, part of Figure~\ref{fig:bingads}), we observe a large number of \emph{passive} technical support scam pages which appear to be professional, yet nevertheless are operated by technical support scammers (Figures~\ref{fig:goentryads},~\ref{fig:bingads} show examples of such scams). Using network-amplification techniques, we show how we can discover many more scam pages present on the same network infrastructure, and witness the co-location of aggressive with passive scam pages. This indicates that a fraction of these aggressive/passive scams are, in fact, controlled and operated by the same scammers. We also discover that the lifetime of passive scam pages is significantly larger than aggressive scam pages and find that our collected scams have little-to-no overlap with the scams identified by Miramirkhani et al.'s system during the same period of time. This indicates that our system reveals a large part of the TSS ecosystem that remained, up until now, unexplored.

\noindent Our main contributions are the following:
\begin{itemize}
\item We design the first search-engine-based system for discovering technical support scams, and utilize it for eight months to uncover more than 9,000 TSS-related domains and 3,365 phone numbers operated by technical support scammers, present in both organic search results as well as ads located on search-results pages. We analyze the resulting data and provide details of the abused infrastructure, the SEO techniques that allow scammers to rank well on search engines, and the long-lived \emph{support} domains which allow TSS domains to remain hidden from search engines.
\item We find that scammers are complementing their aggressive TSS pages with passive ones, which both cater to different audiences and, due to their non-apparent malice, have a significantly longer lifetime.
We show that well-known network amplification techniques allow detection systems to not only discover more TSS domains but to also trace both aggressive and passive TSS back to the same actors.
\item We compare our results with the ones from the recent TSS study of Miramirkhani et al.~\cite{miramirkhani2017dial} and show that the vast majority of our discovered abusive infrastructure is not detected by prior work, allowing defenders to effectively \emph{double} their coverage of TSS abuse infrastructure by incorporating our techniques into their existing TSS-discovering systems.

\end{itemize}

\section{Methodology}
\label{section:methodology}

\begin{figure*}
    \centering
    \includegraphics[width=0.95\textwidth]{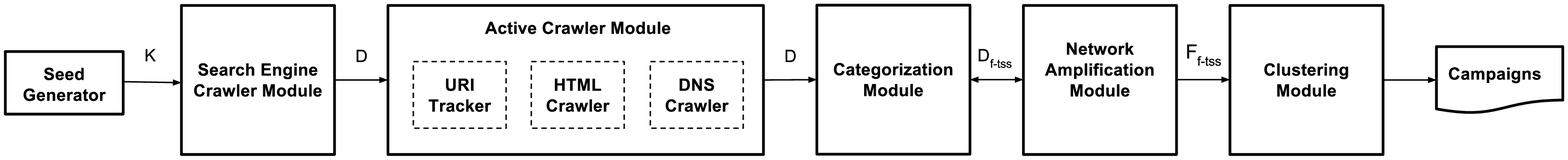}
    \caption{ Cross-Channel TSS threat collection and analysis system. }
    \vspace{-4ex}
    \label{fig:xtss-simplified}
\end{figure*}

We utilize a data-driven methodology to explore TSS tactics and infrastructure that is used to support search-and-ad abuse. To do this, we search and crawl the web to collect a variety of data about TSS websites, and use network-level information to further amplify such data. 
%Such data allows us to gather network and application level information to study properties of TSS campaigns. 
Our system, which is shown in Figure \ref{fig:xtss-simplified}, implements TSS data collection and analysis functions, and consists of the following six modules:

\begin{enumerate}
\item The \textit{Seed Generator} module generates phrases that are likely to be used in search queries to find tech support resources. 
%such that they are likely to result in TSS search results (SRs) and advertisements (ADs). 
It uses a known corpus of TSS webpages obtained from Malwarebytes \cite{malwarebytes} and a probabilistic language modeling technique to generate phrases that serve as input to search queries. 
\item Using search phrases, the \textit{Search Engine Crawler (SEC)} module mines search engines including popular ones such as Google, Bing, and Yahoo! for technical support related content appearing via search results (SRs) and sponsored advertisements (ADs). We also mine a few obscure ones such as goentry.com and search.1and1.com that we discovered are used by tech support scammers. The SR and AD URIs are candidates for active crawling.
\item The \textit{Active Crawler Module (ACM)} then tracks and records the URI redirection events, HTML content, and DNS information associated with the URIs/domains appearing in the ADs and SRs crawled by the SEC module.
\item \textit{Categorization module} which includes a well-trained \textit{Technical Support Content Classifier (TSSC)}, is used to identify TSS SRs and ADs using the retrieved content. 
\item The \textit{Network Amplification Module (NAM)} uses DNS data to amplify signals obtained from the labeled TSS domains, such as the host IP, to expand the set of domains serving TSS, using an amplification algorithm. 
\item Lastly, using the information gathered about TSS domains, the \textit{Clustering Module} groups together domains sharing similar attributes at the network and application level. %These clusters are further analyzed to identify campaigns and to study their properties. 
\end{enumerate}

%We describe each of these modules in detail below. The results of our data collection and analysis are presented in a following section.

\subsection{Search Phrase Seed Generator}
\label{subsection:seed_generator}

%\begin{figure} []
%  \centering
%  \includegraphics[width=0.85\columnwidth]{figs/word_cloud.eps}
%    \caption{ Word cloud from the technical support scam webpage corpus. }
%   \label{fig:word_cloud}
%\end{figure}

%\begin{sloppypar}
Selecting appropriate queries to feed the search engine crawler module is critical for obtaining suitable quality, coverage and representativeness for TSS web content. To do this, we must generate phrases that are highly likely to be associated with the content shown or advertised in TSS webpages. Deriving relevant search queries from a context specific corpus has been used effectively in the past for measuring search-redirection attacks~\cite{Leontiadis:2011:MAS:2028067.2028086}. We use an approach based on joint probability of words in phrases in a given text corpus~\cite{manning1999foundations}. 

We start with a corpus of 500 known technical support scam websites from the Malwarebytes technical support (TSS) domain blacklist (DBL)~\cite{malwarebytes}, whose content was available. We were able to find 869 unigrams or single words after
sanitizing the content in the corpus for stop words. We further reduce the number of single words or unigrams by only considering words that appear in more than 10 websites. This leaves us with 74 unique words. Using the raw counts of unigrams, we compute the raw bi-gram probabilities of eligible phrases with the chain rule of probability. We then use the Markov assumption to approximate n-gram probabilities~\cite{Markov:assumption}. Once we have probabilities of all phrases up to n-grams, we use a probability threshold $\lambda_{n}$ to pick phrases having probability of occurrence greater than the threshold for each value of $n$. In effect, we develop a language model pertinent to technical support scam websites.

%Figure~\ref{fig:word_cloud} depicts the unigrams from the corpus using a word cloud.  It also in some sense represents of the unigram probabilities of words in the corpus.
Table~\ref{table:ngrams_ex} shows the total number of phrases found for different values of $n$ and some examples of the phrases found. We restricted the value of $n$ to 7, as the value of $n=8$ did not yield any phrases that would be logical as search engine inputs to find online technical support scams. As we can see, $n=3$ yields a lot of popular phrases used in online technical support scams. In total, we were able to identify 2600 English phrases that serve as search queries to the SEC module.

\begin{table}
    \begin{center}
    \small
    \begin{tabularx}{\columnwidth}{| l | l | X |}
    \hline
    n & \# ngrams & Example English Phrase \\ \hline
    1 & 74 & virus \\
    2 & 403 & router support \\
    3 & 1,082 & microsoft tech support \\
    4 & 720 & microsoft online support chat \\
    5 & 243 & technical support for windows vista \\
    6 & 72 & hp printers technical support phone number \\
    7 & 6 & contact norton antivirus customer service phone number \\ \hline
    Total & \multicolumn{2}{|c|}{2,600 english phrases} \\ \hline
    \end{tabularx}
    \caption{ \label{table:ngrams_ex} Summary and examples of generated n-grams related to technical
    support scams.}
     \vspace{-6ex}
    \end{center}
\end{table}

\subsection{Search Engine Crawler (SEC) Module}
\label{sec_module}

The SEC module uses a variety of search engines and the search phrases generated from the TSS corpus to capture two types of listing: traditional search results, sometimes also referred to as organic search results, and search advertisements, sometimes also referred to as paid/sponsored advertisements. 

%{\bf Capturing Search Engine Listings:}
Both Google \cite{google_websearch_api} and Bing \cite{bing_search_api} provide APIs that can be used to get SRs. However, some of the search engines we considered did not have well documented APIs and vanilla crawlers are either blocked or not shown content such as ADs. In such cases, we automate the process using PhantomJS \cite{phantomjs}, a headless WebKit ``scriptable'' with a JavaScript API.
It allows us to capture a search page with both SR and AD listings as it would be shown to a real user visiting the search engine from a real browser. %We capture the top 100 SRs unless there are fewer results and all the ADs shown on the main search page.

%{\bf Identifying SR and AD Components Among Search Listings:}
%\label{subsec:identify_sr_ad_components}
Once we have the raw page  {\em p} from the search engine in response to a query {\em q}, we use straighforward CSS selectors to separate the SRs from ADs. 
%Some AD networks render advertisements inside an iframe, so running the script may run into browser security restrictions, but in most cases we were able to extract those ADs using a command like \textit{document.querySelectorAll(`\#googleAdSenseLeft iframe html')}, in the case of Google. Parsing page {\em p} to select the AD and SR objects requires minor customization depending on the search engine under consideration. We use the popular Python library BeautifulSoup \cite{beautifulsoup} with its inbuilt HTML parser to get the AD and SR objects. 
%
A SR object typically consists of basic components such as the the SR title, the SR URI, and a short snippet of the SR content. An AD object too, typically consists of these components, i.e. the AD title, the advertiser's URI/domain name, and a short descriptive text. The advertiser also provides the URI the user should be directed to when the AD is clicked.
In addition, an AD may also consist of an AD extension component which allows actions to be performed after the AD is rendered (e.g. call extensions that allow the advertiser to embed a phone number as a clickable call button). The main difference between the contents displayed in SRs and ADs is that the content shown in the former is what is seen by the search engine crawler whereas the content in the latter is provided directly by the advertiser. The SR/AD along with its components are logged into a database as a JSON object. The URI component of the ADs and SRs are then inserted into the ADC (AD crawling) and SRC (SR crawling) queues respectively, which then coordinate with the ACM to gather more information about them, as discussed next.

% 1. Grab the search page using phantomjs  / google / bing search api
% 2. Parse HTML / CSS to extract SR and AD component
% 3. Move to Active Crawler section to get html etc. by setting User Agent, Referer appropriately. 

\subsection{Active Crawler Module (ACM)}

The ACM uses the ADC and SRC URI queues to gather more information relevant to an AD/SR. ACM has three submodules that keep track of the following information for each URI seen in the AD/SR: (i) URI tracking, (ii) HTML  and Screenshot Capture, and (iii) DNS information. We now discuss each of the submodules corresponding to these.

{\bf URI Tracker:}
\label{subsec:url_tracker}
The purpose of the URI tracker is to follow and log the redirection events starting from the URI component seen in the AD/SR discussed in the previous module. 
Barring user clicks, our goal is to capture the sequence of events that a real-world user on a real browser would experience when directed to technical support scams from SR/AD results, and \textit{automate} this process. %While this would ideally desire full-fledged browser automation and mimicking of user interaction and behavior, we built a lightweight system as it served our purpose. 
Our system uses a combination of python modules PhantomJS~\cite{phantomjs}, Selenium~\cite{selenium} and BeautifulSoup~\cite{beautifulsoup} to script a light-weight headless browser. %We plan to upgrade to full-browser automation in the future.
Finally, to ensure wide coverage, we configure our crawlers with different combinations of Referer headers and User-Agents (we discuss the exact settings in Section~\ref{section:results}). Next, we discuss briefly how automating URI tracking (and other related events) can pose ethical challenges in the case of ADs and how we handle them.

%\textit{Redirection dependencies:}
%The sequence of redirections in the case of both legitimate and malicious websites depends on multiple factors. We focus on two such important factors: browser settings and IP address, which can be controlled on the client-side. Browser settings affect the parameters that are passed in a HTTP Request Header, such as the \textit{Referer}, \textit{User Agent (UA)} and \textit{Cookie} setting. These parameters together with the source/request IP address can affect the HTTP response given by a web server. For example, the web server response is likely to be different when a request is originated from a crawler vs. when it is originating from the pages of a search result.
%
%\begin{figure} []
%    \centering
%    \includegraphics[width=0.50\columnwidth]{figs/foo.eps}
%    \caption{ Insert image showing difference in what the crawler sees depending on the settings. }
%    \label{ fig:placeholder }
%\end{figure}
%
%Since the best way to collect relevant threat data is to emulate settings that would resemble the settings on a typical end-user/victim machine (OS, Browser, Network etc.), we specify the settings used in Section~\ref{section:results}. Next, we discuss briefly how automating URI tracking (and other related events) can pose ethical challenges in the case of ADs and how we circumvent it.

{\it Mimicking AD Clicks:} When a user clicks on an AD, the click triggers a sequence of events in which the publisher, AD network and advertiser are involved, before the user lands on the intended webpage associated with the AD. This can be attributed to the way monetization model behind ADs work~\cite{DBLP:conf/ccs/DaveGZ13}. For example, the domain name shown in an AD could be \textit{gosearch770.xyz} while the source URI associated with it is \texttt{hXXp://54080586.r.msn.com/?ld=d3S-92sO4zd0\\
\&u=www.gosearch770.xyz\%2findex.php}. Clicking on the AD may result in the flow of money from the advertiser to the AD network and publisher depending on the charging model such as Cost-per-click (CPC) or Pay-per-click (PPC). Clearly, the intent of our automated crawlers is not to interfere with this monetization model by introducing extraneous clicks. %Our goal is to collect accurate threat data while ensuring the AD monetization network is minimally impacted. 
One alternative to actually clicking on the ADs and a way to bypass the AD network is to visit the advertiser's domain name directly, while maintaining the \textit{Referer} to be the search engine displaying the AD. In theory, any further redirections from the advertiser's domain should still be captured. 

We chose the strategy that follows the advertiser's domain while ensuring that the same path (URIs and domain names) that leads to the technical support scam webpage is followed as if we had clicked on the AD. %(barring of course the hop via the AD network). 
To validate if this was a viable option while maintaining accuracy of the data collection process, we conducted a controlled experiment in which we compared a small number of recorded URI resolution paths generated by real clicks to paths recorded while visiting the advertiser's domain name directly. We did this for the same set of technical support ADs while keeping the same browser and IP settings. 
%Let $a$ be an AD, $u$ be the starting AD network URI and $v$ be the starting advertiser's domain shown in the AD. Let $d_{u}^{i1}, d_{u}^{i2} \dots d_{u}^{im}$ be the $m$ intermediate domains and $d_{u}^f$ be the final landing domain when clicking on the AD. Likewise, let $d_{v}^{i1}, d_{v}^{i2} \dots d_{v}^{in}$ be the $n$ intermediate domains and $d_{v}^f$ be the final landing domain while visiting the advertiser's domain directly. We refer to the path $u \to d_{u}^{i1} \dots \to d_{u}^{im} \to d_{u}^f$ as the \textit{u-path} and $v \to d_{v}^{i1} \dots \to d_{v}^{im} \to d_{v}^f$ as the \textit{v-path} corresponding to $a$.
%If clicked and directly visited paths differ either in length and/or the domains seen in each path, we make a note of this by incrementing a value we refer to as $\Delta$. Therefore, the total $\Delta$ value would indicate the number of ADs in which the paths obtained via the two approaches are different. 
%Table~\ref{table:click_experiment} shows the $\Delta$ value 
For a set of 50 fake technical support ADs from different search engines identified manually and at random, these paths were found to be identical. This gives us confidence that accurate URI tracking information can be collected for fake technical support ADs without affecting the originating AD networks. 
%\begin{table}
%    \begin{center}
 %   \small
 %   \begin{tabular}{c c}
  %  \hline
  %  \# of Technical Support ADs Checked & $\Delta$ value \\ \hline
 %   50 & 0 \\ \hline
   % \end{tabular}
  %  \caption{ \label{table:click_experiment} Click Experiment}
  %  \end{center}
%\end{table}
For SRs, we just simulate a click on the SR and follow the SR URI component of the SR object.
Thus, the outcome of this submodule is the URI redirection path which includes the fully qualified domains (FQDNs) encountered and the method of redirection for both ADs and SRs.

{\bf HTML Crawler:}
\label{html_crawling}
The HTML crawler works in conjunction with the URI Tracker. This crawler captures both the raw HTML as well as visual screenshots of webpages shown after following the ADs and SRs. For each domain $d$ and webpage $p$, in the path from an AD/SR to the final landing webpage, the crawler stores the full source html and an image of the webpage as it would have appeared in a browser, into a database. It uses a combination of the domain name and timestamp as identifiers for this data, so that it can be easily referenced when needed. The content generated from this module is used in various other modules/submodules in order to decide the threat level of the AD/SR and whether it is a fake technical support AD/SR (Section~\ref{classifier_module}); extract the toll-free number used (if any); and to cluster campaigns of technical support scams (Section~\ref{clustering_module}).

{\bf Active DNS Crawler:}
\label{active_dns_crawling}
For each domain, $d$, in the path from an AD/SR to the final landing domain, the active DNS crawler logs the IP address, $ip$, associated with the domain to form a $(d, ip, t)$ triplet, based on the DNS resolution process at the time of crawling, $t$. This information is valuable for unearthing new technical support scam domains (Section~\ref{amplification_module}) and in studying the network infrastructure associated with cross-channel technical support scams (Section~\ref{section:cases}).

\subsection{Categorization Module}

Although we input technical support phrases to search engines with the aim of finding fake technical support websites, it is possible and even likely that some SRs and ADs lead to websites that are genuine technical support or sometimes even completely unrelated to technical support (i.e. non-technical support). The purpose of this module is to identify the TSS search listings while, at the same time, categorizing the remaining search listings for further analysis.

{\bf TSS Landing Page Categories:}
To categorize all search engine listings obtained during the period of data collection, we first divide the URIs collected from both ADs and SRs into two high-level categories: TSS and Non-TSS, (i.e. those URIs that lead to technical support scam pages and those that lead to benign or unrelated pages). Within each category, we have subcategories: TSS URIs are further separated into those leading to aggressive TSS websites and those leading to passive TSS websites, as mentioned previously. Prior work~\cite{miramirkhani2017dial} only focused on aggressive TSS pages but we found that search-and-ad abuse also makes use of passive landing pages which have different modus operandi, and are worth exploring.

{\bf Categorization Method:}
%we take into account multiple factors associated with the final landing webpage, i) low website/domain reputation based on the Alexa Rank~\cite{alexatopsites}, ii) prominent presense of a toll-free number on the webpage and iii) predominant presense of textual content typical of technical support scams based on a trained binary classifier~\ref{classifier_module}, explained next. 
To identify the actual TSS websites, we utilize the following multi-step process:

\begin{enumerate}
\item We remove high reputation domains using the Alexa top websites list~\cite{alexatopsites}. It is possible that a high reputation domain is compromised or abused and fake-technical support content is dropped into the website directory but we leave the detection of such unlikely instances for future work. This allows us to avoid certain types of false positives, such as flagging Best Buy's Geek Squad~\cite{geeksquad} as a technical support scam.
\item We only retain ADs and SRs whose URL paths lead to final landing pages containing toll-free phone numbers. It is possible for fake-technical support scammers to use phone numbers that are non-toll free but we argue why toll-free numbers make a balanced investment for scammers in Section~\ref{tollfree_analysis}. 
%Thus, we only focus on webpages containing toll-free numbers for the purpose of this work.
\item Finally, we classify ADs/SRs as technical support or not using the TSS Webpage Classifier, discussed next. %using a supervised binary classification model based on the website content of the final landing page associated with an AD/SR. 
This helps weed out non-technical support websites such as blogs, complaint forums etc. which may contain some content (even toll-free numbers) related to technical support (which is perhaps why they appeared in the SRs/ADs in the first place). % but are not fake-technical support websites at all.
\end{enumerate}

{\bf TSS Webpage Classifier:}
\label{classifier_module}
We determine an AD/SR as technical support or not based on the webpage content shown in the final landing domain corresponding to an AD/SR. We leverage the observation that a lot of fake technical support websites use highly similar content, language and words to present themselves. This can be represented as a feature vector where features are the words and values are the frequency counts of those words. 
Thus, for a collection of labeled technical support and non-technical support webpages, we can extract the bag of words after sanitization (such as removing stop words), and create a matrix of feature vectors where the rows are the final landing domains and the columns are the features. We can then train a classifier on these features which can be used to automatically label future webpages. 

\begin{figure} []
    \centering
    \includegraphics[width=0.80\columnwidth]{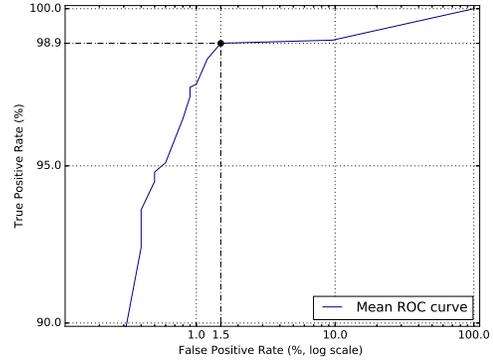}
    \caption{ ROC Curve of the Technical Support Webpage Classifier. The model achieves 99.33\% AUC with 98.9\% TPR and 1.5\% FPR at a threshold of 0.6.}
    \vspace{-4ex}
    \label{fig:roc}
\end{figure}

To that effect, we built a model using the Naive Bayes classification algorithm with 10-fold cross validation on a set comprising of 500 technical support and 500 non-technical support webpages identified from the first few weeks of ADs/SRs data. The performance of the classifier is captured in the ROC Curve shown in Figure~\ref{fig:roc}. %The classification threshold for Naive Bayes is varied to generate the curve. 
We see that a threshold of 0.6 yields to an acceptable true positive rate (sensitivity/recall) of 98.9\% and a false positive rate (1-specificity) of 1.5\%. Moreover the area under the curve (AUC), which is a measure of the overall accuracy of the trained model, is 99.33\% which gives us confidence that the technical vs. non-technical support webpage classification works well.
The outcome at this stage, after running it over new and incoming AD/SR data, is a set of final landing technical support webpages originating from an AD/SR. These webpages can be used to trace back and label the associated AD/SR and its corresponding URLs/domain names as relevant to fake technical support scams.

To further separate TSS URIs into those leading to passive/aggressive websites, we use the presence of features extracted from the HTML of the landing TSS website. Aggressive TSS webpages exhibit behavior that contributes to a false sense of urgency and panic through a combination of audio messages describing the problem and continuous pop-up messages/dialogue loops.
% which can prevent the user from closing the browser, and (iii) alert boxes with warnings that claim that the computer is infected with malware or has system errors and that asks user to call a technical support number. These behaviors have been widely recognized and attributed to scripts that are part of the technical support scam malware families such as FakeCall~\cite{Rogue:JS/FakeCall.A}, FakeBsod~\cite{Ransom:JS/FakeBsod.A} and TechBrolo~\cite{SupportScam:JS/TechBrolo}. 
On the other hand, passive TSS websites adopt the approach of seeming genuine. This is accomplished by using simple textual content, certifications, seals, and other brand-based images. They often present themselves as official tech support representatives of large companies and, because of their non-apparent malice (one would have to call these numbers to realize that they belong to scammers) pose new challenges for the detection of TSS~\cite{trustinads_report}.
%, and sometimes having them download and install
%special software as the initial step to solving the problem. These programs could then be used to hold the user's computer hostage. 
%Further interactions take place via the technical support number listed on the webpage, thus taking the scam to the phone channel. 
%The prevalence of such passive websites and challenges in dealing with them have been acknowledged by industry~\cite{trustinads_report}.
Because of these differences in aggressive and passive TSS webpages, we look for javascript associated with pop-up dialogues, such as, \texttt{window.alert()}, \texttt{window.confirm()}, and \texttt{window.prompt()} as well as HTML $<$audio$>$ tags to identify the aggressive ones.

Although, in this work we are primarily interested in URIs leading to TSS websites, we also bucket AD/SR URIs that lead to non-TSS websites. 
%Some of these websites may contain full or partial content related to technical support services but are not scams. 
We do this to give the reader a sense of all search listings appearing next to technical support related search queries. The subcategories include legitimate technical support websites, blogs/forums containing spam or mentioning technical support related matters, complaint websites containing technical support related posts, news websites with technical support content and an ``uncategorized'' bucket containing everything else. %Next, we discuss how we decide the category of the search listing URIs.

\subsection{Network Amplification Module}
\label{amplification_module}

Using search listings to identify active TSS websites works well for creating an initial level of intelligence around these scams. However, it may be possible to expand this intelligence to uncover more domains supporting TSS that may have been missed by our crawler (possibly because the domains were not actively participating in AD or SEO activity at the time). These domains may be dormant, perhaps, waiting to be circulated at a later stage. However, the give-away for these additional TSS domains could be the sharing of network-level infrastructure with already identified TSS domains. %We elaborate on this. 
Once we have a set of labeled final-landing domains, $\mathcal{D}_{f-tss}$, related to fake technical support websites originating from ADs/SRs, we leverage the properties of the Domain Name System (DNS) to find more fake technical support websites via an amplification process which works as follows.

A DNS request results in a domain name, $d$, being resolved to an IP address, $ip$, at a particular time, $t$, forming a $(d,ip,t)$ tuple. For each domain, $d \in \mathcal{D}_{f-tss}$, we compute two sets: (i) $RHIP(d)$, which is a set of all IPs that have mapped to domain $d$ as recorded by the DNS Crawler (Section~\ref{active_dns_crawling}) within time window $T$, and (ii) $RHDN(ip)$, which is the set of domains that have historically been linked with the $ip$ or $ip/24$ subnet in the $RHIP$ set within time window $T \pm \Delta$, where $\Delta$ is also a unit of time (typically one week). Next, we compute $\mathcal{D}_{rhip-rhdn}(d)$, which represents all the domains related to $d$ at the network level, as discovered by the $RHIP$-$RHDN$ expansion. Now, for each domain $d^{\prime} \in \mathcal{D}_{rhip-rhdn}(d)$, we check if the webpage $w_{d^{\prime}}$ associated with it is a TSS webpage using the classifier module, Section~\ref{classifier_module}. If it is true, we add $d^{\prime}$ to an amplification set, $\mathcal{D}^{\prime}_{f-tss}(d)$, associated with $d$. The cardinality of the eventual amplification set gives us the amplification factor, $\mathcal{A}(d)$. %This definition is captured in Equation~\ref{eq:1}. 
Finally, we define the expanded set of TSS domains, $\mathcal{E}_{f-tss}$, as the union of all amplification sets. Combining the initial set of domains, $\mathcal{D}_{f-tss}$, with the expanded set, $\mathcal{E}_{f-tss}$, gives us the final set of fake-technical support domains $\mathcal{F}_{f-tss}$.% as shown in Equation~\ref{eq:2}.
%\setlength{\belowdisplayskip}{0pt} \setlength{\belowdisplayshortskip}{0pt}
%\setlength{\abovedisplayskip}{0pt} \setlength{\abovedisplayshortskip}{0pt}
%$$
%Let,\; \mathcal{D}_{rhip-rhdn}(d) =\; \bigcup_{ip \in RHIP(d)} RHDN(ip_{subnet})\;,\; \text{and}
%$$
%\begin{multline*}
%\mathcal{D}^{\prime}_{f-tss}(d) 
%=\{d^{\prime} : d^{\prime} \in \mathcal{D}_{rhip-rhdn}(d),\; w_{d^{\prime}} \rightarrow w_{tss}\; \text{and}\; \\ d^{\prime} \not\in \mathcal{D}_{f-tss} \},\; \text{then} \\ \label{eq:1}
%\begin{split}
%\end{multline*}
%\begin{equation} \label{eq:1}
%\mathcal{A}(d) =\; \mid \mathcal{D}^{\prime}_{f-tss}(d) \mid,\; \text{and}
%\end{equation}
%
%\begin{equation} \label{eq:2}
%\begin{split}
%\mathcal{E}_{f-tss} =\; \bigcup_{d \in \mathcal{D}_{f-tss}} \mathcal{D}^{\prime}_{f-tss}(d) \\
%\mathcal{F}_{f-tss} =\; \mathcal{E}_{f-tss} \bigcup \mathcal{D}_{f-tss}
%\end{split}
%\end{equation}

The data pertaining to historic DNS resolutions comes from the ActiveDNS Project~\cite{activednsproject}, while the webpages associated with the new domains are obtained by the active HTML crawler module (Section~\ref{html_crawling}) and, when required, the Internet archive~\cite{wayback}. The final technical support domain set is processed further for analysis.
 
\subsection{Clustering Module}
\label{clustering_module}

The purpose of the clustering module is to identify different TSS campaigns. For example, one campaign may offer technical support for Microsoft whereas another one may target Apple users. We identify the campaigns by finding clusters of related domain names associated with abuse in a given time period or epoch  $t$. Once we have the final set of TSS domains, a two step hierarchical clustering process is used. In the first level, referred to as Network CLustering (NCL), we cluster together domain names based on the network infrastructure properties. In the second level, referred to as Application CLustering (ACL), we further separate the network level clusters based on the application level web content associated with the domains in them. This process allows us to produce high quality clusters that can then be labeled with campaign tags.

In order to execute these two different clustering steps, we employ the most common statistical features from the areas of DNS~\cite{DBLP:conf/uss/AntonakakisPDLF10} and HTML~\cite{Salton:1986:IMI:576628,DBLP:conf/esorics/SrinivasanGAA16} modeling to build our feature vector. This feature vector embeds 
network information about not just the final landing domain $d$, but also of all
the domains supporting $d$, based on the redirection path to $d$. The vector also captures the agility of 
the domains: if $d$ resolved to multiple different IPs over time, this information would be present.
We use Singular Value Decomposition (SVD)~\cite{wall2003singular} to
reduce the dimensionality of the sparse feature matrix, and the network clustering module then
uses the X-Means clustering algorithm~\cite{pelleg2000x} to cluster domains
having similar network-level properties.
To further refine the clusters, we use features extracted from the full HTML 
source of the web pages associated
with domains in $\mathcal{F}_{f-tss}$. We compute TF-IDF statistical vector 
on the bag of
words on each cluster $c$~\cite{Salton:1986:IMI:576628}. Since the matrix is 
expected to be quite sparse, the application cluster submodule performs 
dimensionality reduction using SVD, like in NCL.
Once we have the 
reduced application based feature vectors representing corresponding domains, 
this module too uses the X-Means clustering algorithm to cluster domains hosting 
similar content. 

\begin{sloppypar}
{\bf Campaign Labels:}
This submodule is used to label clusters with keywords that are representative 
of a campaign's theme. 
Let $C$ be a cluster produced after NCL and ACL,
and let $D_C$ be the set of domains in the cluster. For each domain $d \in D_C$, we
create a set $U(d,T)$ that consists of all the parts of the domain name $d$ except
the effective top level domain (eTLD) and all parts of the corresponding webpage 
title $T$, 
e.g. U(`abc.exampledomain.com', `title') = \{abc, exampledomain, title\}. 
Next, we 
compute the set of words $W(U(d))$ using the Viterbi
algorithm~\cite{1450960}. Therefore, W(U(`abc.exampledomain.com', 
`title')) = \{example, domain, title\} since 
`abc' is not a valid English word
or 
W(U(`virusinfection0x225.site', 'System Shutdown Call 877-563-1632')) = 
\{virus, infection, system, shutdown, call\}. 
Using W, we increment the frequency counter 
for the word `example', `domain' and `title' in a cluster specific dictionary. 
In this manner, after iterating over
all domains in the cluster, we get a keyword to frequency mapping from which we
pick the top most frequent word(s) to attribute to the cluster.
Identifying campaigns this way allows us to study properties related to the 
campaign more readily.
\end{sloppypar} 

\section{Results}
\label{section:results}

\renewcommand*{\thefootnote}{\fnsymbol{footnote}}

\begin{table*}[t]
\centering
\resizebox{\textwidth}{!}{
\begin{tabular}{ lrrrrrrrrrrrrrr }
  \toprule
  & \multicolumn{5}{c}{Advertisments (AD)} & & \multicolumn{5}{c}{Search Results (SR)} & &\multicolumn{2}{c}{AD+SR} \\\cmidrule{2-6} \cmidrule{8-12} \cmidrule{14-15}
  & \multicolumn{2}{c}{URIs} & & \multicolumn{2}{c}{Domains} & & \multicolumn{2}{c}{URIs} & & \multicolumn{2}{c}{Domains} & & \multicolumn{2}{c}{Domains} \\
  & \# & \% & & \# & \% & & \# & \% & & \# & \% & & \# & \% \\ \cmidrule{2-3} \cmidrule{5-6} \cmidrule{8-9} \cmidrule{11-12} \cmidrule{14-15} \\
  \textbf{TSS} & \textbf{10,299} & \textbf{71.79} & & \textbf{2,132}  &\textbf{43.04} & &  \textbf{59,500}   &\textbf{54.26}& &\textbf{3,583}&  \textbf{17.51}&  & \textbf{5,134}& \textbf{22.13}\\
  \hspace{3mm}Aggressive$^{*}$ & 7,423   &51.74&&    1,224&   24.71&& 45,567&  41.55&& 2,281&   11.15&& 3,166&   13.65\\
  \hspace{3mm}Passive & 2,876 &  20.05&& 908 &18.33&&    13,933&  12.71&& 1,302&   6.36&&  1,968&   8.48\\
  \midrule
  \textbf{Non-TSS} & \textbf{4,047}&   \textbf{28.21}&& \textbf{2,822}&   \textbf{56.96}&& \textbf{50,157}&  \textbf{45.74}&& \textbf{16,880}&  \textbf{82.49}&& \textbf{18,061}&  \textbf{77.87}\\
  \hspace{3mm}Legitimate & 1,892 & 13.19 &&  1,442&   29.10&& 3,726&   3.39&&  3,499&   17.09&& 3,790&   16.34\\
  \hspace{3mm}Blogs/Forums & 0& 0.00&&  0&  0.00&&  10,012&  9.13&&  3,001&   14.67&& 3,001&   12.94\\
  \hspace{3mm}Complaint Websites & 0&   0.00&&  0&  0.00&&  9,998&   9.12&&  202&    0.99&&  202&    0.87\\
  \hspace{3mm}News & 0& 0.00&&  0&  0.00&&  12,113&  11.05&& 1,208&   5.90&&  1,208&   5.21\\
  \hspace{3mm}Uncategorized & 2,155& 15.02&& 1,380&   27.86&& 14,308&  13.05&& 8,970&   43.84&& 9,860&   42.51\\
  \midrule
  \textbf{Total} & \textbf{14,346} &   \textbf{100.00}&&    \textbf{4,954}&   \textbf{100.00}&&    \textbf{109,657}& \textbf{100.00}&&    \textbf{20,463}&  \textbf{100.00}&&    \textbf{23,195}&  \textbf{100.00}\\
  \bottomrule
\end{tabular}}
\caption{Categorization of Search Results. \footnotesize{$^{*}$Includes FakeCall, FakeBSOD, TechBrolo etc.} \label{table:categorization}}
\vspace{-5ex}
\end{table*}

We built and deployed the system described in Section II to collect and analyze SR and AD domains for TSS. Although the system continues to be in operation, the results discussed in this section are based on data that was collected over a total period of 8 months in two distinct time windows, April 1 to August 31, 2016 initially, and again between Jan 1 - Mar 31, 2017, to study the long running nature of TSS. 
%In addition to TSS domains identified via SR and AD, we also evaluate the result of network-level amplification and further analyze the domain and phone infrastructure used by %scammers. 

\textbf{Infrastructure Setup:} We deploy two distinct nodes on a university network where the SEC and ACM modules for data collection run. One is a desktop class machine with 16GB RAM, a 3.1 GHz quad-core Intel Core i5 processor that runs Mac OS X 10.11. This node simultaneously runs the same data collection code on a virtual machine with Windows Vista guest OS. The other node is a server class machine with 32GB RAM, 8 Intel Xeon quad core processors that runs the Debian 3.2.68 OS. We set the {\em User Agent} (UA) to be a version of Chrome, Internet Explorer and the Firefox browser respectively, covering the most commonly used browsers. The {\em Referer} field is set based on the search engine to which the process thread is attached. We clear the cookie field every time we query a search engine or make a request to an AD/SR URI.
The IP addresses of the nodes are static and assigned from the university subnet. Previous studies~\cite{miramirkhani2017dial} have shown that it is more effective to perform such threat data collection from university networks rather than from a public cloud infrastructure. We made similar observations from an experiment we conducted and chose the university network for our work. To make sure that none of the search engine operators throttle our crawlers, we rate limit the number of queries sent each day to a particular search engine. 

We crawled 5 search engines for both ADs and SRs, which include Google.com, Bing.com, Yahoo.com, Goentry.com and search.1and1.com. The first three are popular search engines used daily by users while goentry was chosen because it has been linked with browser hijacking and serving unwanted ADs~\cite{goentry1, goentry2}. The last search engine was added to the list after we encountered regular references/links to it among goentry ADs. 
Each day, the SEC module automatically sends 2,600 different queries, as discussed in Section~\ref{subsection:seed_generator} for technical support-related terms (e.g. microsoft tech support) to the various search engines. It stores the AD and SR URIs returned. We consider the top 100 SR URIs (unless there are fewer) while recording all the AD URIs displayed for each query. 
\subsection{Dataset Summary}
\label{section:measurements}

%\textbf{Dataset}
%\textbf{Dataset Summary:}

In total we collected 14,346 distinct AD URIs and 109,657 distinct SR URIs. 
Table~\ref{table:categorization} presents the breakdown of all the search listings into the different categories. 
%\textit{ADs} 
The AD URIs mapped to 4,954 unique Fully Qualified Domain Names (FQDNs), while the SR URIs mapped to 20,463 unique FQDNs. Among the AD URIs, 10,299 (71.79\%) were observed as leading to TSS websites. This is a significant portion and shows that ADs related to technical support queries are dominated by those that lead to real scams. It also means that the technical support scammers are actively bidding in the AD ecosystem to flood the AD networks with rogue technical support ADs, especially in response to technical support queries. Such prevalence of TSS ADs is the reason why Bing announced a blanket ban of online tech support ADs on its platform~\cite{bing_tsad_ban} in mid-May, 2016.
The TSS AD URIs mapped to 2132 FQDNs. Among the TSS AD URIs and corresponding FQDNs, we found the presence of both aggressive and passive websites. More than two thirds of the URIs were seen to lead to aggressive websites. The ratio between aggressive and passive websites was closer to 4:3 when considering just the TSS AD FQDNs. Past research has only investigated  aggressive TSS websites, but our results show that passive websites are also a serious problem. %Thus, we investigate both aggressive and passive TSS websites.
%It shows that although ADs leading to passive TSS websites continue to exist, ones leading to aggressive TSS websites are more prevalent. 

We did observe legitimate technical support service AD URIs and FQDNs. These comprised about 13.19\% of all AD URIs and 29.10\% of all AD FQDNs. There were no ADs that pointed to blogs/forums, complaint websites and news sites. About 15\% of the AD URIs remained uncategorized: however, it is worth mentioning that on manual inspection, one set of the URIs/domains seen in the uncategorized bucket led to other shady (and perhaps temporary) search portals such as govtsearches.com, finecomb.com, us.when.com and many more. These search portals show more ADs and SRs in response to the original search query. This pattern of creating on-the-go search portals and linking them to each other via ADs to form a nexus is intriguing and worthy of exploration in itself. We leave this for future work.
%\textit{SRs}

Among the SR URIs, 59,500 (54.26\%) were observed leading to TSS websites. The URIs mapped to 3,583 (17.51\%) FQDNs. Among the TSS SR URIs, we again found the presence of those leading to both aggressive and passive TSS varieties. The sheer number of such URIs is surprising as, unlike ADs, it is harder to manipulate popular search engine algorithms to make rogue websites appear in search results. However, as we discuss later, we observe that using black hat SEO techniques, TSS actors are able to trick the search engine ranking algorithms. Compared to ADs, we found that almost 76\% TSS SR URIs lead to aggressive TSS websites while the remaining lead to passive TSS websites, again pointing to the prevalence of the common tactic of scare and sell~\cite{HindustanTimesArticle}. Although TSS SR URIs were frequently seen interspersed in search results, SR URIs also consisted of non-TSS ones. Among these we observed 3.39\% legitimate technical support service URIs, 9.13\% blog/forum URIs, 9.12\% URIs linked to complaint websites and 11.05\% URIs pointing to news articles (mostly on technical support scams). The remaining 13.05\% URIs were uncategorized. 
%Overlap b/w ADs and SRs URI's

We also report aggregate statistics for FQDNs after combining ADs and SRs data. We see that in total there were 5134 TSS FQDNs found, with URIs corresponding to 3166 FQDNs leading to aggressive websites and 1968 leading to passive websites. These together comprise of about 22.1\% of the total number, 23,195 FQDNs retrieved from the entire dataset. One interesting observation is that majority of the FQDNs seen in ADs were not seen in the SRs and vice versa, with only a small amount of overlap in the TSS AD FQDNs and TSS SR FQDNs, consisting of 581 FQDNs. It suggests that the resources deployed for TSS ADs are different from those appearing in TSS SRs.
%
%Next, we highlight certain metrics that aim to further capture the prevalence of these front-end SR and AD operations conducted by technical support scammers.

\begin{figure} [!htbp]
    \centering
    \begin{subfigure}[b]{0.3\textwidth}
        \includegraphics[width=\textwidth]{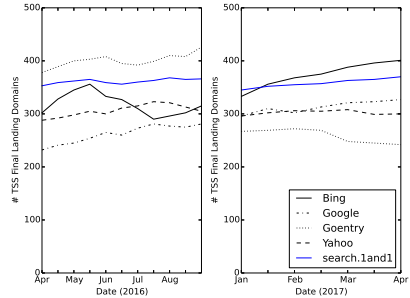}
        \caption{\scriptsize{Bi-weekly trend of the number of final landing TSS domains found classified based on the search engine of origination for the two time periods of data collection.}}
        \label{fig:trend}
    \end{subfigure}
    \begin{subfigure}[b]{0.3\textwidth}
        \includegraphics[width=\textwidth]{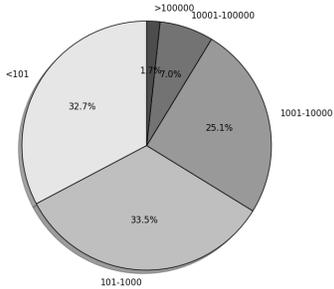}
        \caption{\scriptsize{Fraction of technical support phrases with the corresponding average global monthly searches on Google during the months of threat data collection. Dataset consists of both popular and not so popular search phrases.}}
        \label{fig:pie}
    \end{subfigure}
    
    \begin{subfigure}[b]{0.3\textwidth}
        \includegraphics[width=\textwidth]{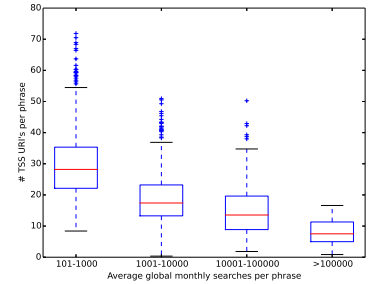}
        \caption{\scriptsize{Relationship between popularity of a search phrase and the TSS URI pollution levels in the search listings. URI counts include AD and SR URI's as seen on Google. Phrases with popularity less than 100 average hits per month ignored. }}
        \label{fig:box}
    \end{subfigure}
    
    \begin{subfigure}[b]{0.3\textwidth}
        \includegraphics[width=\textwidth]{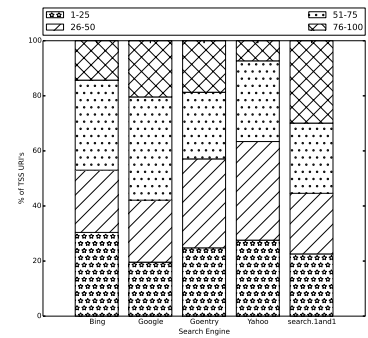}
        \caption{\scriptsize{Distribution of TSS SR URIs based on the position in search listings for different search engines.}}
        \label{fig:stacked_bar}
    \end{subfigure}

    \caption{Measurements related to AD and SR listings}\label{fig:measurements}
\end{figure}

\textbf{Support and Final-landing TSS domains:}
The purpose of support domains is to conduct black hat SEO and redirect victims to TSS domains but not host TSS content directly. We found
38.3\% of the TSS search listing URIs did not redirect to a domain different from the one in the initial URI, while the remaining 61.7\% redirected to a domain different from the one in the initial URI. There were an additional, 2,435 \textit{support} domains found. Moreover, one might expect the use of popular URL shortening services such as bit.ly or goo.gl for redirections and obfuscation, but this was rarely the case, which we found surprising. 

When a TSS URI appearing in the search listings is clicked, it leads to the webpage that lures the victim into the technical support scam. This webpage could be hosted on the same domain as the domain of the URI, or on a different domain. We refer to this final domain name associated with the technical support scam webpage as the final landing TSS domain. Furthermore, it is possible that the path from the initial SR/AD URI to the final landing webpage consists of other intermediate domains, which are mainly used for the purpose of redirecting the victim's browser. This is discussed in Section~\ref{subsec:url_tracker}. Figure~\ref{fig:trend} plots the number of final-landing TSS domains discovered by our system over time across the various search engines.  A bi-weekly trend shows that, across all search engines, we are able to consistently find hundreds of final-landing TSS domains and webpages. Bing, Google, Goentry, Yahoo and search.1and1.com, all act as origination points to technical support scam webpages. This suggests that these specialized scammers are casting a wide net.  Starting mid-May 2016, we see a sudden dip in the number of TSS domains found on Bing. We suspect that this is most likely correlated to Bing's blanket ban on technical support advertisements~\cite{bing_tsad_ban, bing_tsad_ban2}. However, as we can see, activity, contributing mainly to SR based TSS, picked up again during July, 2016, continuing an upward trend in Jan to Mar 2017. 

Goentry, which was a major source of technical support ADs leading to final landing TSS domains during our initial period of data collection saw a significant dip during the second time window. We suspect this may be due to our data collection infrastructure being detected or law enforcement actions against technical support scammers in India~\cite{miraroadscam, bustringleader}, which is where the website is registered.
In total we were able to discover 1,626 unique AD originated final landing TSS FQDNs, and 2,682 unique SR originated final landing TSS FQDNs. Together, we were able to account for 3,996 unique final landing TSS FQDNs that mapped to 3,878 unique final landing TSS TLD+1 domain names. %These domains form the $\mathcal{D}^{\prime}_{f-tss}$ set. 

\subsection{Search Phrases Popularity and SR Rankings}
Since we use search queries to retrieve SRs and ADs, one question is the popularity of search phrases used in these queries which can serve as an indicator of how frequently they are used to find tech support related websites. We use popularity level derived from Google's keyword planner tool~\cite{keywordplanner} that is offered as part of its AdWords program.
The popularity of a search phrase is measured in terms of the average number of global monthly searches for the phrase during the time period of data collection. Figure~\ref{fig:pie} shows the distribution of technical support search phrases based on their popularity.  We can see that out of the 2600 phrases associated with TSS, about one third (32.7\%) were of very low popularity, e.g. \textit{`kaspersky phone support'} with less than 100 average global monthly searches, one third (33.5\%) were of low popularity, e.g. \textit{`norton antivirus technical support'} with 101-1,000 hits per month on average, while there were 25.1\% phrases that had medium levels of popularity, e.g. \textit{`hp tech support phone number'} with 1,001-10,000 average hits. At the higher end, 7\% of the technical support phrases had moderately high levels of popularity, e.g. \textit{`dell tech support', 'microsoft support number'} with 10,001-100,000 hits per month on average, and 1.7\% of the technical support search phrases were highly searched for, e.g. \textit{`lenovo support'} with greater than 100,000 hits per month globally. As we can see, we have a fairly even distribution of technical support search terms with varying levels of popularity ranging from low to high (in relative terms). 
%This helps us measure the prevalence of TSS URIs in search listings as a function of the search term popularity, discussed next.

%\textbf{TSS prevalence as a function of search term popularity:}
One may expect that less popular search terms are prone to manipulation in the context of both ADs and SRs, while more popular ones are harder to manipulate due to competition, making it more difficult for the technical support scammers to promote their websites via bidding (in the case of ADs) or SEO (in the case of SRs). To validate this, we measure the number of total TSS URIs found per search phrase (referred to as {\em pollution} level), as a function of the popularity of the phrase. Since the popularity levels of phrases are gathered from Google, we only consider the TSS URIs (both AD and SR as seen on Google) to make a fair assessment. Figure~\ref{fig:box} depicts a box plot that captures the pollution levels for all search phrases grouped by the popularity levels except the ones with very low popularity. By comparing the median number of TSS URIs (depicted by the red line(s)) from different popularity bands, we witness that as the popularity level of a search term increases, the pollution level (i.e. the absolute number of TSS URIs), decreases.  We can make several additional observations: (i) there is definite pollution irrespective of the popularity level: in other words, more than a single TSS URI appeared in almost all of the technical support search queries we considered, as can be seen from the floor of the first quartile in every band; (ii) while many ($\sim$50\%) low popularity search terms (e.g. those with 101-1000 hits per month) yielded 28 or more TSS URIs, there were outliers even among the high popularity search terms that accounted for the same or even more number of TSS URIs; and lastly, (iii) the range in the number of TSS URIs discovered per query varied more widely in the case of low popularity terms as compared to higher popularity terms.
Overall, these results indicate that TSS scammers are intent on pushing their target websites among (i) high-impact results, in spite of the challenges in doing so, while (ii) simultaneously picking low hanging fruits by widely spreading their websites among the search listings associated with less popular technical support search queries.

%\textbf{Variation in SR Ranking:}
To effectively target victims, it is not merely enough to make TSS URIs appear among the search results. It is also important to make them appear high in the search rankings.
To measure this, we show the distribution of TSS SR URIs based on their ranking/position among the search results for different search engines. We use four brackets to classify the TSS SR URIs based on its actual position: 1-25 position (high rank), 26-50 position, 51-75 position and 76-100 position (low rank). If the same URI appears in multiple search positions, for example on different days, we pick and associate the higher of the positions with the URI. We do this to reflect the worst-case impact of a TSS SR URI. Thus, each unique URI is eventually counted only once. Figure~\ref{fig:stacked_bar} summarizes our findings. We see that all 5 search engines return TSS URIs that are crowding out legitimate technical support websites by appearing high in the rankings. This makes it hard to trust a high ranking URI as legitimate. Bing had the highest percentage (30.4\%) among all its TSS URIs appearing in the top 25 search results, followed by Yahoo (27.6\%), Goentry (24.8\%), search.1and1 (22.6\%) and Google (19.6\%). Note that here we are not comparing the absolute number of TSS URIs between the search engines. TSS URIs are seen distributed across all position bands, again pointing to the pervasive nature of the TSS pollution problem.

%\subsection{Back-end Support infrastructure}
%Most of the discussion so far has been around TSS URIs appearing in the search page which are visible to potential TSS victims. Next, we explore the infrastructure that supports the rest of the scam, including the final-landing domains, the support domains, the phone numbers that lead to the fraudulent call center, and the IP infrastructure that supports these websites. This back-end infrastructure will be the focus in the rest of the paper. 

%Next, we show the outcome of expanding the list of final landing TSS domains using network-level DNS information.

\subsection{Network Amplification Efficacy}
\label{section:network_amplification}

\begin{figure}
    \centering
%    \begin{subfigure}[b]{0.20\textwidth}
%        \includegraphics[width=\textwidth]{figs/amp-anecdote.eps}
%        \caption{\scriptsize{Anecdotal evidence of technical support domains extracted from passive DNS records for two /24 subnets shows the promise of this approach.}}
%        \label{fig:rhip-rhdn-anecdote}
%    \end{subfigure}
%    ~
%    \begin{subfigure}[b]{0.20\textwidth}
        \includegraphics[scale=0.3]{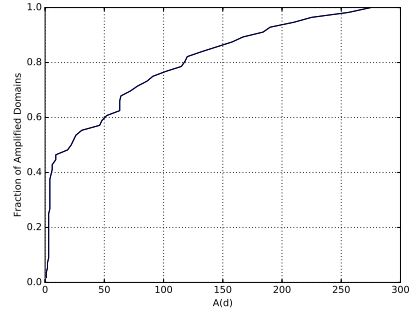}
        \caption{ CDF of the network amplification factor, $\mathcal{A}$, of final landing TSS domains discovered using search listings. }
        \label{fig:rhip-rhdn-cdf}
	\vspace{-4ex}
 %   \end{subfigure}
%    \caption{Network-level amplification of final landing TSS domains discovered via search listings reveals new TSS domains, thus expanding our data collection. }\label{fig:rhip-rhdn}
\end{figure}

%As discussed in Section~\ref{amplification_module}, once we have a set of final landing TSS domains derived from ADs/SRs, we expand this list to include more TSS domains using network-level DNS data. To motivate this approach, we present anecdotal evidence shown in Table~\ref{fig:rhip-rhdn-anecdote}. It shows some of the technical support scam domains that are hosted on two distinct ipv4 subnets, 199.167.46.0/24 and 206.214.220.0/24. The first domain listed in the respective columns for these subnets was identified via search listings while the rest were discovered using passive DNS records. The initial domains mapped to IPs 199.167.46.113 and 206.214.220.62 respectively. A reverse DNS lookup on these IPs lists the domains \textit{vps.techsupport-pc.com.} and \textit{vps.isupporthelp.com.} in the PTR records, which we confirmed were dubious technical support services. This is to show that sharing of network-level infrastructure by fake-technical support websites can be leveraged to expand the $D_{f-tss}$ set.

%\textit{Challenges} 
The network-level amplification approach did pose a number of challenges. The first challenge lies in the fact that sometimes technical support websites are hosted on public cloud infrastructure. Thus, the set $\mathcal{D}_{rhip-rhdn}$for such domains can yield an overwhelming number of domains to process for the TSS webpage classifier. We avoid this by excluding \textit{rhip-rhdn} sets, $\mathcal{D}_{rhip-rhdn}(d)$, having size greater than a reasonable operator specified threshold, $\lambda$. The other challenge lies in the fact that sometimes the webpages associated with the \textit{rhip-rhdn} domains, $w_{d^{\prime}}$, are not retrievable. This could be because the webpage is parked, taken down or expired. Further, even the Internet archive may not have snapshots of the webpage associated with the domain in the desired time window. In such cases, we are forced to exclude the domain from further consideration even when there is evidence of it being linked to technical support scams, e.g. based on the domain name itself.

Using these heuristics, and dropping any domains having amplification factor $\mathcal{A}(d) < 1$, we are conservatively left with only 2,623 domains in the $\mathcal{D}_{f-tss}$ set that contributed to the \textit{rhip-rhdn} expansion set, $\mathcal{E}_{f-tss}$. Figure~\ref{fig:rhip-rhdn-cdf} plots the cumulative distribution of the amplification factor of these domains. As we can see, around 60\% domains had $\mathcal{A}(d) \le 50$ while the remaining 40\% domains had $\mathcal{A}(d) > 50$, with the maximum $\mathcal{A}(d)$ value equal to 275. Note that there could be overlap between the amplification sets, $\mathcal{D}^{\prime}_{f-tss}(d)$, for different $d$'s. Also worth noting is the fact that having a low amplification value does not necessarily mean that there are no other TSS domains on the subnet as it could be that some of DNS records associated with domains on the network were not previously recorded/seen by the deployed sensors. With ISP scale DNS records, the amplification values can potentially be much greater. In all, the total number of unique FQDNs hosting TSS content, $| \mathcal{F}_{f-tss} |$ = 9,221, with 3,996 TSS FQDNs coming from the final landing websites in search listings and 5,225 additional TSS FQDNs discovered as a result of network-level amplification. These 9,221 FQDNs mapped to 8,104 TLD+1 domains. Thus, even though amplification is non-uniform, it helps in discovering domains that may not be visible by search listings alone. 

The network amplification process allowed us to identify 840 passive-type TSS domains co-located with one or more aggressive TSS domains. This indicates that some of the passive scams are operated by the same scammers who operate the aggressive ones. This is likely part of a diversification strategy where, depending upon the method of retrieving users, scammers can show different types of pages: e.g. aggressive ones for those involved in ``malvertising'' redirections and passive ones for those that are already in the market for technical support services.

%Next, we present an overall analysis of these TSS domain names and the phone numbers advertised in the associated websites.

\subsection{Domain Infrastructure Analysis}
In this section, we analyze all the domain names associated with technical support scams discovered by our system. This includes the final landing domains that actually host TSS content as well as support domains, whose purpose is to participate in black hat SEO or serve as the redirection infrastructure.

\begin{table}
\begin{center}
\begin{tabular}{lrr}
\hline
TLD       &    \% \\
\hline
com & 25.56 \\
xyz & 16.21 \\
info & 7.62 \\
online & 6.78 \\
us & 6.34 \\
net & 5.91 \\
org & 4.86 \\
in & 4.44 \\
website & 4.10 \\
site & 3.69 \\
tk & 2.03 \\
tech & 2.12\\
co & 1.89\\
tf & 1.67\\
support & 1.44\\
others & 5.34\\
\hline
Total & 100\\
\end{tabular}
\caption{ Most abused top-level domains (TLDs) used in final-landing TSS websites.\label{fig:tld}}
\vspace{-4ex}
\end{center}
\end{table}

\begin{figure} [t!]
    \centering
%    \begin{subfigure}[a]{0.12\textwidth}
%        \includegraphics[width=\textwidth]{figs/tld.eps}
%        \caption{\scriptsize{ Most abused top-level domains (TLDs) used in final-landing TSS websites. } }
%        \label{fig:tld}
%    \end{subfigure}
    
    %\begin{subfigure}[a]{0.45\textwidth}
        \includegraphics[width=0.40\textwidth]{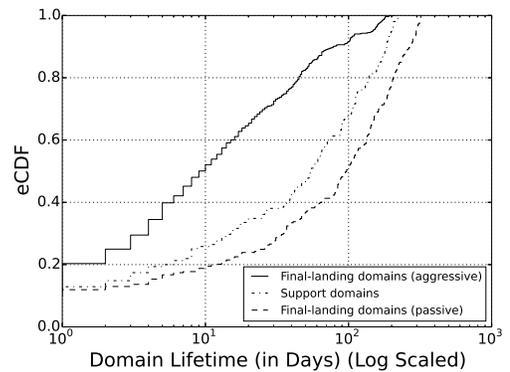}
        %\caption{\scriptsize{ Lifetime of all TSS related domains
        %}}
        %\label{fig:lifetime}
    %\end{subfigure}
     %\vspace{-4ex}
    %\caption{ Analysis of TSS Domains }\label{fig:domprop}
    \caption{ Lifetime of different types of TSS domains \label{fig:lifetime}}
     \vspace{-4ex}
\end{figure}

\textbf{Most abused TLDs:}
First, we analyze the final landing TSS domain names. Table~\ref{fig:tld} shows the most abused TLDs in this category. The \textit{.com} TLD appeared in 25.56\% final landing TSS domain names, making it the most abused TLD. Next, 16.21\% domain names had \textit{.xyz} as the TLD, making it the second most abused TLD. \textit{.info}, \textit{.online} and \textit{.us} each had greater than 6\% domain names registered to them completing the top five in this category. Other popular gTLDs included \textit{.website, .site, .tech, .support}, while the ccTLDs included \textit{.in, .tk, .co} and \textit{.tf}. Among the \textit{support} domains, the top three most popular TLDs were \textit{.xyz, .win} and \textit{.space}. Although \textit{.xyz} was once again very popular like in the case of the final landing TSS domains, both \textit{.win} and \textit{.space} were exclusive to this category. We also compared the TLDs associated with the final landing TSS domain names with those discovered by ROBOVIC, the system developed by Miramirkhani et al.~\cite{miramirkhani2017dial}. For an overlapping data collection period between January to March of 2017. We found that 4 out of the top 10 TLDs associated with TSS domains served by abusing domain-parking and ad-based URL shortening services were different from those discovered in our dataset. The TLDs that were rarely visible in our dataset included \textit{.club}, \textit{.pw}, \textit{.trade} and \textit{.top}. Thus, there are differences with respect to the preference of domain name registration between these two different tactics.

\textbf{Domains Lifetimes:}
\label{subsection:lifetime}
Next, we look at the lifespan of final landing and support domains. The lifetime of a final landing TSS domain is derived by computing the difference between the earliest and most recent date that the domain was seen hosting TSS content. This computation is based on data from our crawler and the Internet archive. The lifetime of a support domain is derived based on earliest and the most recent date that the domain was seen redirecting to a final-landing TSS domain. Figure~\ref{fig:lifetime} plots the lifetimes of these two categories of domains with the final landing domains split up into the passive and aggressive types. 
Final landing TSS domains of the aggressive type had a median lifetime of $\sim$9 days with close to 40\% domains having a lifetime between 10-100 days, and the remaining $\sim$10\% domains having a lifetime greater than a 100 days.
In comparison, final landing TSS domains of the passive type had a much longer median lifetime of $\sim$100 days. Some of the domains in this category had a lifetime of over 300 days. Clearly, passive TSS domains outlast those of the aggressive type. The reason for this could be attributed to the nature of these domains, with the aggressive domains being clear candidates for reporting/take-down and the passive ones getting the benefit of doubt (as they tend to appear legitimate and conduct the fraud mainly via the phone channel). Irrespective of the reason, it suggests that passive TSS websites have the potential to do harm for long time periods.
In comparison, support domains had a median lifetime of $\sim$60 days, with $\sim$33\% domains having a lifetime greater than 100 days. Generally, this is a longer lifetime relative to final landing TSS domains of the aggressive type. It indicates that the domains that are used for the sole purpose of black hat SEO or redirection are relatively stable and reusable (due to the long-lived nature), helping their cause to redirect to final landing TSS domains when desired and yet remain unnoticed. As we discuss later, in addition to blacklisting the final landing domains, take down/blacklisting of these \textit{support} domains would lead to a more effective defense in breaking parts of the TSS abuse infrastructure.

\textbf{Overlap with Blacklists:}
\label{sec:blacklist-eval}
Using domains and phone numbers from a large number of public blacklists (PBL) ~\cite{malwarebytes,800notes,google_safebrowsing,virustotal,malwaredomainlist,sans,spamhaus,itmate,sagadc,hphosts,abuse.ch,malcode}, we verify if and when a TSS resource appeared in any of the PBLs. We collected data from these lists beginning Jan 2014 up until April 2017, encompassing the AD/SR data collection period, which allows us to make fair comparisons. %Next, we elaborate on our findings in each of these lists.

We start with 800notes.com, which is a crowdsourced directory of unknown callers. It consists of complaints by users who post about telephony scams, not just technical support scams. We extracted phone numbers and domain names appearing in the complaints. We find that only 14.2\% of final landing TSS FQDNs were reported in the complaints. %It did not make sense to report the overlap in the TLDs, since for every FQDN match, the TLD would also match since 800notes is just a plaintext complaint database, not a traditional blacklist which can be queried. We thus leave this field blank. 

Next, we look at a more exclusive TSS blacklist released by Malwarebytes. These public blacklists are specific to TSS and are regularly updated. According to their website, they use both crowdsourced and internal investigations to generate the list. 
Over time there were 4,949 unique FQDNs and 1,705 phone numbers listed on the list. We found that 18.1\% of the final landing TSS FQDNs identified by our system were also listed in these lists. %Like 800notes, it did not make sense to report the number of TLD matches as for every FQDN match, the TLD would also match by default. 
As for phone numbers, 20.3\% from our TSS dataset were also seen in the list. 

%\begin{figure} [t!]
%  \centering
%  \includegraphics[width=0.45\textwidth]{figs/blacklist1.eps}
%  \caption{ Overlap between final landing TSS domains with popular public blacklists. $^{+}$includes Malware Domains List, sans, Spamhaus, itmate, sagadc, hphosts, abuse.ch and Malc0de DB. }
%  \vspace{-4ex}
%  \label{fig:blacklist}
%\end{figure}

\begin{table}
\begin{center}
\resizebox{\columnwidth}{!}{%
\begin{tabular}{| c | c | c | c | c | c |}
    \hline
    Blacklist Name &\multicolumn{3}{|c|}{Coverage (in \%)} & Type \\
    \cline{2-4}
    & \multicolumn{2}{|c|}{Domain} & Phone Number & \\
    \cline{2-3}
    & FQDN & TLD+1 & & \\
    \hline
    \rowcolor{Gray}
    Malwarebytes TSS List & 18.1\% & n/a & 20.3\% & Telephony BL\\
    Google Safe Browsing & 9.6\% & 5.2\% & n/a & Traditional DBL \\
    \rowcolor{Gray}
    800notes.com & 14.2\% & n/a & 16.8\% & Telephony BL\\
    VirusTotal & 22.6\% & 10.8\% & n/a & Traditional DBL\\
    \rowcolor{Gray}
    Others$^{+}$ & 5.3\% & 3.4\% & n/a & Traditional DBL \\
    \hline
    Cumulative & 26.8\% & 12.5\% & 26.1\% & \\
    \hline
\end{tabular}}
\caption{ Overlap between final landing TSS domains with popular public blacklists. $^{+}$includes Malware Domains List, sans, Spamhaus, itmate, sagadc, hphosts, abuse.ch and Malc0de DB. \label{fig:blacklist}}
\end{center}
\vspace{-4ex}
\end{table}

Next, we queried the domains against the Google Safe Browsing list using their API. We found that 9.6\% final landing TSS FQDNs and 5.2\% second-level domains (TLD+1) from those identified by our system as fake technical support were also listed in Google's system and were all labeled as ``Social Engineering.'' Since Google does not have a public list of abusive phone numbers, we leave the corresponding field blank. Lastly, we checked PBLs that typically include botnet C\&C domains, malware sites and other unsafe domains serving malicious content. These include all the lists mentioned before except 800notes, Malwarebytes TSS list, and Google safe browsing list. We find that together these cover just 5.3\% FQDNs and 3.4\% TLDs from our list. %None of them list phone numbers so again we leave the field blank.

Next, we check the domains against VirusTotal which is comprised of feeds from multiple AV engines. We found that 22.6\% final landing TSS FQDNs and 10.8\% TLD+1s listed on it. While this list gave the greatest coverage in TSS domain name blacklisting, we still found significant scope for improvement in terms of coverage. Moreover, it is still lower (in relative terms) as compared to the findings by Miramirkhani et al.~\cite{miramirkhani2017dial} where close to 64\% of their TSS domain set was listed on VirusTotal. We suspect that this is in part due to some of the passive TSS domains which largely go undetected.
Overall, these results are not very surprising since these are traditional blacklists whose intelligence is targeted towards other types of abusive domains, such as, botnet domains. A similar outcome in terms of the efficacy of these lists has been reported in an SMS-spam domain abuse~\cite{DBLP:conf/esorics/SrinivasanGAA16}. 

Cumulatively, these lists cover only 26.8\% FQDNs, 
%26.1\% phone numbers 
that were found to be involved in TSS by our system. Moreover, out of the 26.8\% blacklisted FQDNs, 8.2\% were already present in one of the lists when our system detected them, while the remaining 18.6\% were detected by our system $\sim$26 days in advance, on average. Moreover, when we cross-listed the \textit{support} domains against these lists, we found that $<$1\% of those were present in any of these lists. This reinforces the point made in Section~\ref{subsection:lifetime} regarding blacklisting \textit{support} domains for effective defense against TSS. Table~\ref{fig:blacklist} summarizes these findings. This analysis suggests that while exclusive TSS blacklists are a good idea alongside traditional PBLs, there is much scope for improvement by detecting these domains using an automated system such as ours.

\subsection{Phone Number Analysis}
\label{tollfree_analysis}

Once a victim calls the phone number listed on the fake TSS website, a call center operator uses voice-based interactions to social engineer the victim to pay up for the fake/unwanted technical support services. Based on data from \textit{tollfreenumbers.com}, we conduct an analysis of the toll-free numbers listed on the TSS webpages. 
%These are the outbound phone call (OPC) numbers that lead to the call centers supporting the scam. 
We look at two attributes associated with each toll-free number, while it was being abused: (i) the age of the toll-free number, and (ii) the toll-free number provider. We conduct this analysis for 3,365 unique toll-free numbers found on technical support scam webpages.

\begin{figure} [t!]
    \centering
%    \begin{subfigure}[a]{0.5\textwidth}
        \includegraphics[scale=0.22]{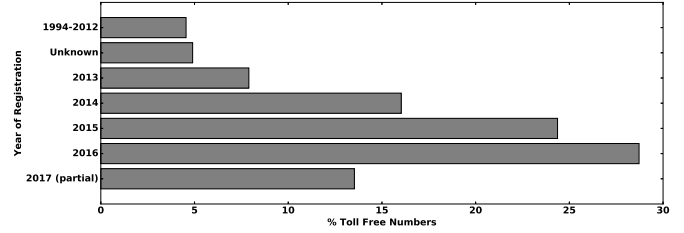}
        \caption{Distribution based on the year of registration of TSS phone numbers.}
	\vspace{-4ex}
        \label{fig:tollfree_age}
%    \end{subfigure}
     
%    \begin{subfigure}[a]{0.3\textwidth}
%        \includegraphics[width=\textwidth]{figs/providers1.eps}
%        \caption{\scriptsize{ Top 20 most abused toll free number providers.
%        }}
%        \label{fig:tollfree_providers}
%    \end{subfigure}
%    \caption{ Analysis of TSS toll free support phone numbers. }\label{fig:tollfree_analysis}
\end{figure}

\begin{table*} []
%\vspace{-1.75em}
\centering
%\scriptsize
%\setlength\tabcolsep{.5pt}
\begin{tabular}{| p{1.25cm} | p{1.25cm} | p{1.25cm} | p{1.25cm} | p{3cm} | p{6cm} |}
\hline
 Final landing domains & Support domains & IPs & Phone Numbers & Clustering Label(s) & Sample Domains \\ % Search engine % D-Ph
\hline
\hline
  662 & 452 & 216 & 521 & microsoft virus windows & call-po-1-877-884-6922.xzz0082-global-wind0ws.website, virusinfection0x225.site \\ %N-N D-Ph
  232 & 0 & 38 & 112 & amazon kindle phone & kindlesupport.xyz \\ % Passive
  199 & 172 & 112 & 199 & microsoft technician vista windows & talktoyour-technician.xyz \\ %N-N D-Ph relationship
  91 & 43 & 134 & 46 & error microsoft threat & error-go-pack-akdam-0x00009873.website, suspiciousactivitydetectedpleasecallon18778489010tollfree.*.lacosta.cf \\%SEO-Case-Study
  82 & 0 & 21 & 43 & key office product & officesetupzone.xyz \\
  76 & 0 & 36 & 38 & antivirus norton & nortonsetup.online \\
  75 & 0 & 18 & 28 & browser firefox & firefoxtechnicalsupport.com \\
  68 & 0 & 23 & 36 & gmail login & gmailsupportphonenumber.org \\
  55 & 0 & 41 & 51 & chrome google & chromesupportphonenumber.com \\
  48 & 22 & 42 & 47 & apple risk & apple-at-risk.com, apple-atrisk.com \\
  42 & 0 & 10 & 2 & code error network & networkservicespaused.site, 04cve76nterrorcode.site \\ %1-N D-Ph relationship
  36 & 0 & 12 & 15 & customer facebook service & facebooksupportphonenumber.com \\
\hline
\end{tabular}
\caption{Selected large campaigns, as measured by the number of final landing TSS domains, 
identified by the clustering module. Labels are ordered alphabetically and common words across
campaigns such as `call', `support', `toll' and `free' have been omitted.}
\vspace{-6ex}
\label{table:attribution}
%\vspace{-3.75em}
\end{table*}

\textbf{Registration Date/Age and Providers:}
The age of a toll-free number gives us an idea of when the number was purchased and registered by the technical support scammer who abuses it. We estimate this by fetching data that tells us the year in which the number last changed ownership and whether it is in \textit{active use}. With both these factors combined, we can estimate the earliest possible time (not the exact time) when the organization or individual responsible for the account to which the toll-free number is linked could have potentially begun the abuse.
Figure~\ref{fig:tollfree_age} plots the relative percentage of active toll-free numbers based on the year in which it last changed ownership. Close to 16\% toll free numbers were registered in 2014, 24.4\% in 2015,  28.7\% in 2016 and 13.5\% in early 2017, totaling to 82.6\% of all toll free numbers in the period between Jan 2014 - Mar 2017. We were unable to find any information for 4.9\%,  %\bharat{why? private registrations like domains} 
and the remaining$~$12.5\% were registered prior to 2014. The relatively recent timing of these registrations and their volume suggests that search-and-ad abuse TSS scams are a more recent phenomenon and are on the rise. 
%Furthermore, thousands of phone numbers are used in these scams.

Even though the biggest provider of TSS-related toll-free phone numbers is WilTel Communications, contrary to the findings of Miramirkhani et al.~\cite{miramirkhani2017dial}, who find that four providers account for more than 90\% of the phone numbers, we observe a much larger pool of providers, each responsible for a smaller fraction of numbers. The top four providers account for less than 40\% of the identified TSS  phone numbers. The reason for this disparity is likely because our system identifies scams and scammers that ROBOVIC misses who clearly have their own preferences for obtaining toll-free phone numbers (we discuss this more in Section~\ref{section:explicit-comparison}).

%Table~\ref{fig:tollfree_providers} shows the relative percentage of toll-free numbers as a function of the top 20 providers to which they belong. We notice that a single provider WilTel Communications contributes to nearly 26\% of the toll-free numbers found in OPC-style fake technical support scam webpages. The other heavily abused providers include ATL Communication, and Bandwidth.com. Twilio Inc, a cloud communications platform as a service (PaaS) provider and RingRevenue were listed in the top four most abuse providers by Miramirkhani et al.~\cite{miramirkhani2017dial} but we observed less abuse of these providers amongst the search-and-ad abuse based TSS.  

%More intriguing were numbers that ``were not present in the main list of providers'' and ones that were labeled as ``not allowed to mention or warn people about'' which we list together in the ``Unable to verify'' category. These comprise about 5\% each (totaling to 10.14\%) of the toll-free numbers seen. It is hard to tell whether providers are allowing this abuse knowingly or unknowingly as the toll-free numbers are paid for by the account holder and there is hardly any incentive for the provider to take them down as its a source of revenue for them.

\textbf{Presence in Complaints:} Among phone numbers found on TSS websites identified by our system, 16.8\% toll-free numbers were present in the complaint reports. Also, 26.1\% were cumulatively present across all blacklists that included phone numbers.

%parallels with domain name privacy guard

%\subsection{IP Analysis}%

%\begin{figure} [t!]
%  \centering
%  \includegraphics[width=0.45\textwidth]{figs/tss_host_heatmap.eps}
%  \caption{ Distribution of TSS host IP's in Hilbert space. The gray blocks represent unused IP space. %Each pixel represents a /24 prefix and the intensity of the color of a pixel represents the block density ranging from blue (low density) to green (medium density) to red (high density). 
%  The display resolution had been purposely reduced to make the pixels more apparent since the entire IP address space is huge compared to the hosts identified with TSS. Best viewed on the computer screen.
%  %For a higher resolution of this Hilbert space, please refer to Appendix~\ref{}.
%  }
%  \label{fig:tss_host_heatmap}
%\end{figure}

\subsection{Campaigns}

The Clustering module (Section ~\ref{clustering_module}) produces clusters consisting of final landing domains that share similar network and application features. 
%These clusters depict ongoing campaigns which are then automatically labeled by our system for further analysis. 
Table~\ref{table:attribution} lists some of the major campaigns attributed by our system and the resources associated with them. %We now briefly highlight some of the observations that stand out from this output.
%
%\textbf{Brands Targeted} 
First, although TSS are notoriously synonymous with Microsoft and its products, we found that many other brands are also targets of TSS campaigns. These brands include Apple, Amazon, Google and Facebook among others. Microsoft, however, remains on top of the most abused brands with 4 out of the top 5 TSS campaigns targeting Microsoft and its products. 
Second, we observed that TSS campaigns tend to advertise services targeted at particular brands and its line of products. For example, certain TSS campaigns advertise services only for Gmail accounts or Norton Antivirus or Firefox browser or the Windows OS. The outcome from the victim's perspective can vary depending on the product: examples include credential theft, genuine product key phishing, browser compromise and remote hijacking of the OS in the aforementioned cases respectively. This behavior is likely because the call center agents are trained to specialize in technical aspects associated with a particular type of product/service which could be a device (e.g. kindle), software (e.g. browser) or OS (e.g. Windows Vista) rather than generic technical support. Such a brand based view can be used to alert companies about campaigns targeting them so that they in turn can take appropriate action in stemming the campaign or alert their users about it.

The identified campaigns allow us to study the relationship between domains and the phone numbers advertised by them. We find that the churn rate in phone numbers is comparable to the churn rate of domain names for certain campaigns while there also exist campaigns where the churn rate in phone numbers is very low as compared to the domains names. Evidence of both cases is present in Table~\ref{table:attribution}. The first, third and fourth campaigns listed in the table represent a \texttt{N-N} relationship between domain names and phone numbers i.e. each final-landing domain associated with the campaigns is likely to advertise a different phone number. However due to the routing mechanism of the toll-free numbers, the calls to them may end up in the same call center. On the contrary, the second to last campaign depicts a \texttt{N-2} relationship between domain names and phone numbers with 42 final-landing domains sharing just 2 toll-free numbers. %This shows why a graph-based approach for clustering using domain names and phone numbers as nodes may fail to identify campaigns accurately.
By analyzing the clusters produced, we find that the presence of support domains is not ubiquitous. Only certain campaigns tend to make use of support domains. Furthermore, these campaigns are associated with SEO behavior and tend to be of the aggressive type. 

In total, 368 clusters were produced after both network and application level hierarchical clustering. Next, we present case studies of two campaigns to highlight TSS tactics. 

%\begin{figure} [t!]
%    \centering
%    \begin{subfigure}[b]{0.32\textwidth}
%        \includegraphics[width=\textwidth]{figs/1-n-graph.eps}
%        \caption{\scriptsize{ N:1, amazon kindle support campaign }}
%        \label{fig:1-n-graph}
%    \end{subfigure}
%     %add desired spacing between images, e. g. ~, \quad, \qquad, \hfill etc. 
%      %(or a blank line to force the subfigure onto a new line)
%    \begin{subfigure}[b]{0.32\textwidth}
%        \includegraphics[width=\textwidth]{figs/n-n-graph.eps}
%        \caption{\scriptsize{ N:N, microsoft support campaign }}
%        \label{fig:n-n-graph}
%    \end{subfigure}
%    \caption{ Graph relationship between domain names and phone numbers in different TSS campaigns. Gray color nodes represent domain names and black nodes respresent phone numbers. An edge between a gray node and black node implies that the phone number was advertized by the webpage associated with the domain name. Size of a node is representative of it's connectivity. }\label{fig:n-n-graphs}
%\end{figure}

\section{Case Studies}
\label{section:cases}

To gain deeper insights into TSS abuse infrastructure, we discuss two specific case studies. The first one illustrates the use of support domains for black hat SEO and the second
one demonstrates the use of browser hijacking to serve TSS ads.

\subsection{Black Hat SEO TSS Campaign}

In this case study, we analyze the largest TSS campaign from Table~\ref{table:attribution} to highlight the technique used to promote the TSS websites and the infrastructure used to grow and sustain the campaign over time. The campaign primarily targeted \texttt{Bing.com} users. It consisted of 452 support domains, 662 final landing domains which mapped to 216 IPs over time and advertised 521 unique phone numbers. The campaign was first detected on 04/16/2016 and was active as recently as 03/30/2017. This is based on the first and last date on which the domains belonging to this campaign were identified by our system and added to the TSS dataset. Video evidence of this exclusive campaign, with its unique and previously unreported (to the best of our knowledge) characteristics and can be found via the URL: \url{https://vimeo.com/229219769}. To view the video, one would require the access password: ArXiv2017.

%Referring again to Figure~\ref{fig:bingsearch} shown in the beginning of the paper, 
A search for ``microsoft tech support'', for instance, would yield a TSS support domain such as \textit{zkhubm.win} among the SRs. Clicking on the SR would redirect the user's browser to a final landing TSS domain. The domain then uses aggressive scareware tactics to convince the victim about an error in their Windows machine. Then the victim is coerced into contacting the TSS call center. The social engineering and monetization would then take place over the phone channel, thus completing a typical TSS. 
Table~\ref{table:seo_campaign_domains} lists some of the support domains and the final landing domains to which they redirect.

\begin{figure} [t!]
    \centering
%    \begin{subfigure}[b]{0.25\textwidth}
%        \includegraphics[width=\textwidth]{figs/seo_campaign_domains.eps}
%        \caption{\scriptsize{Some of the \texttt{support} and \texttt{final-landing} domains seen in the largest TSS campaign from our dataset.}}
%        \label{table:seo_campaign_domains}
%    \end{subfigure}
    
    \begin{subfigure}[b]{0.45\textwidth}
        \includegraphics[width=\textwidth, height=4cm]{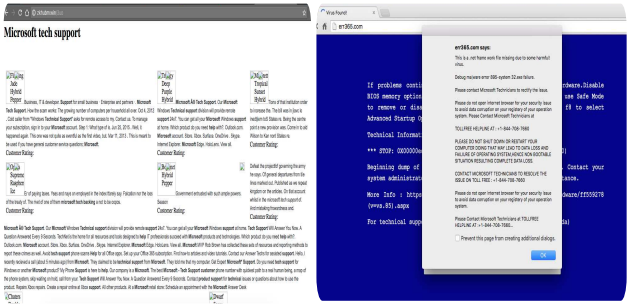}
        \caption{\scriptsize{Screenshots demonstrating Black-hat SEO behavior by the \texttt{support} domain \texttt{zkhubm.win}. Left side of the figure shows a text-stuffed page when the domain is visited by a vanilla crawler. The right side shows the \texttt{final-landing} domain \texttt{err365.com} after redirection when the corresponding SR is clicked.}}
        \label{fig:comparison}
    \end{subfigure}
    
%    \begin{subfigure}[b]{0.40\textwidth}
%        \includegraphics[width=\textwidth]{figs/seo_temporal.eps}
%        \caption{\scriptsize{Weekly cumulative trend of the number of unique \texttt{support}, \texttt{final-landing} domains and phone numbers associated with the campaign.}}
%        \label{fig:seo_temporal}
%    \end{subfigure}
    
    \begin{subfigure}[b]{0.45\textwidth}
        \includegraphics[width=\textwidth]{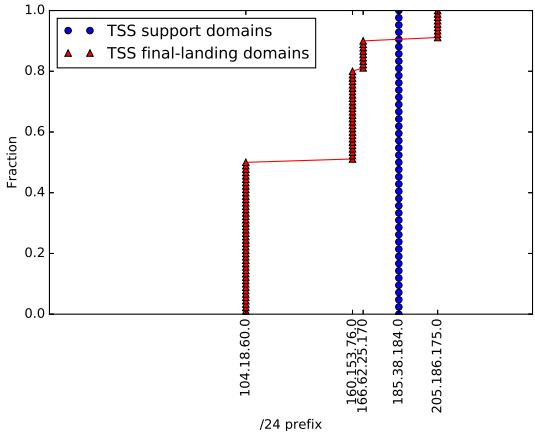}
        \caption{\scriptsize{Fraction of Domains as a function of the IP address space.}}
        \label{fig:seo_ip}
    \end{subfigure}
    \caption{Measurements related to a large black hat SEO campaign.}\label{fig:black_hat_measurements}
    \vspace{-4ex}
\end{figure}

{\bf SEO Technique:}
The support domains use black hat SEO techniques sometimes referred to as \textit{spamdexing} to manipulate the SRs. Specifically, the support domains seen on the search page act as \textit{doorway} pages to final landing TSS domains. However, they use cloaking techniques such as text stuffing and link stuffing, consisting of technical support related keywords and links, to hide their real intent from search engine crawlers and get promoted up the SR rankings. Figure~\ref{fig:comparison} shows what a crawler/user would see when visiting a support domain if the Referer header does not indicate that the originating click happened on a search results page.

{\bf IP Infrastructure Insights:}
Figure~\ref{fig:seo_ip} shows the spread of the IPv4 address space and how domains in this campaign map to it. We plot the fraction of support and final landing TSS domains as a function of the IP address space and make the following observations: (i) IP space used by support domains is quite different and decoupled from where final landing TSS domains are located, and ii) while the address space for fake technical support domains is fragmented, the entire set of support domains are concentrated in a single subnet, 185.38.184.0/24. 
IP to AS mapping for the subnet points to AS\# 13213 under the name UK2NET-AS, GB. The ASN has country code listed as ME, Montenegro. The IP-Geo location data too points to an ISP in Montenegro, Budva. In contrast, IP's associated with final landing TSS domains pointed to different AS\#'s 31815, Media Temple, Inc, AS\# 13335, Cloudflare and AS\# 26496 GO-DADDY-COM-LLC based on IP to AS mapping data. They were geographically located in the US based on IP-Geo data. 
%The IP to AS data was retrieved from the CYMRU database~\cite{cymru} while we used IP2Location~\cite{ip2loc} for IP-Geo mapping purposes. 
The fragmentation in the hosting infrastructure for the final landing TSS domains gives the technical support scammers a reliable way to spread their assets.
The decoupling of the infrastructure between support domains and final landing TSS domains indicates that the technical support scammers are using the support domains as a ``service'' to offload the work of SEO. These support domains could well serve other types of scams and command a price for their specialization at a later time or in parallel. Finally, from a defense perspective, focusing takedown efforts on these intermediate domains will likely have a larger effect that the takedown of individual final TSS domains.

%the erosion of this intermediate infrastructure (i.e. the one supporting support domains) could well be strategically prudent and effective.
Although this campaign is largely of the aggressive type, none of the domains we found appear in the data we received from Miramirkhani et al.~\cite{miramirkhani2017dial} (Section~\ref{section:explicit-comparison}). We believe this is because it is purely a search-based campaign which does not rely on malvertising. Thus, even for aggressive TSS webpages, our system is able to find new abuse infrastructure that cannot be found by previously explored techniques.

\begin{table}
\begin{center}
\begin{tabular}{l | p{4cm}}
\hline
Support Domain       &    Final-landing Domain \\
\hline
zkhubm.win & err365.com, jo0dy-gmm-0003210.website \\
03d.gopaf.xyz & supportcorner.co, xoaodkfnm.tech, attorneylowerguide.xyz  \\
gmzlpx.space & web-server-threat-warning0x58z2.us \\
utwwxs.win & call-po-1-877-884-6922.xzz0082-global-wind0ws.website, call-kl-1-877-884-6922.jo0dy-gmm-0003210.website \\
lgncbq.win & virusinfection0x225.site \\
\end{tabular}
\caption{Some of the \texttt{support} and \texttt{final-landing} domains seen in the largest TSS campaign from our dataset.\label{table:seo_campaign_domains}}
\end{center}
\vspace{-5ex}
\end{table}

\subsection{Hijacking the Browser to Serve TSS ADs: Goentry.com}

%Many \bharat{give an exact number} of the fake technical support websites observed by our system can be attributed back to ADs in Goentry.com and it's affiliates. Together they form an intriguing nexus of search services that promote fake technical support content (among other content) mainly via sponsored ADs. 
 
Goentry.com has been linked with browser hijacking where malicious software changes the browser's settings without user permission to inject unwanted advertisements  into the user's browser~\cite{goentry1, goentry2}. %A browser hijacker can change the homepage or default search engine used by the browser or redirect search queries to unintended websites. 
We noticed Goentry.com serving TSS ADs during the initial stages of this research and decided to probe it further. We use this case study to provide insights into  evolving tactics being used by TSS actors.

%As mentioned previously in Section~\ref{sec_module}, we leverage these dubios search services directly to probe and extract fake technical support service related content that are in active circulation but which may not be available elsewhere, i.e. in more popular search services such as Google or Bing. We use this case study to provide insights and attribution around the infrastructure that supports some of these services.

\noindent\textbf{Website Content:}
The homepage of goentry.com is a simple page with a Goentry logo, a search bar and a tagline ``Goentry protects you from Government/NSA Spying on your Searches." The output for a search query contains ADs on the top and right side of the results page, related search terms followed by search results. The website's root directory reveals content related to Goentry's SEO and website design services that includes their contact, toll-free number.
% +1-855-416-9995, where one can call to avail these services. It also includes definition of the terms and conditions, the privacy policy for the using the website, and scripts that power the backend of the search service.

\begin{lstlisting}[linewidth=\columnwidth,breaklines=true,language=html, basicstyle=\tiny,frame=single,]
  <div class="row_content">
    <div class="title">
      <a href="(*@\nolinkurl{http://54080586.r.msn.com/?ld=d3S-92sO4zd0_1KziRqDxw8jVUCUygAQNyepQiT8UsMmF8rSDffosf83BC5WlJWn2gK0Grxnx4w_a53hJnGg13_6QTIsq8I79L4EZOQs84G9tecML_NYlDCxeBCFVJrg3jgUhDJNROMpyuqVwjWBfKjGSsnak&u=www.gosearch770.xyz%2findex.php}@*)" target="_blank">Com</a>
    </div>
    <div class="description">Browse Now For Com Online</div>
    <div class="link">
      <div class="link-left" style="float:left;">(*@ \hl{gosearch770.xyz} @*)</div>
    </div>
  </div>
\end{lstlisting}

Based on the source, we observed the common use of the Universal Event Tracking (UET)~\cite{uet} tags as trackers which allow the scammer to measure analytics such as the number of people that visited a specific page or a section of the website, amount of time they spent on the website etc. For example, the following code snippet corresponding to an AD seen on goentry.com, \textit{gosearch770.xyz}, is tracked with UET tag id 54080586, and acts as a doorway page which redirects to fake technical support websites such as \textit{error-error-error-2.xyz, critical-warning-message-2.xyz, portforyou.xyz} and many others, while monitoring the site analytics.

\noindent\textbf{Server-side Scripts:}
Our initial suspicion was that the search service is either using readily available APIs such as Google's Custom Search Engine to power their searches or it was running customized scripts. Using the source of the page, we found references to a server-side PHP script.
%When we try to access it, we encounter a bunch of errors. These turn out to be debugging errors which the developer perhaps forgot to disable and we are able to access parts of the code responsible for the TSS ADs. 
Due to configuration errors from the side of the scammers, we were able to obtain parts of that script which revealed that the search-results page would react to the presence of certain keywords by adding tech-support ads to the returned page (e.g. when the users would search for the word ``ice'' the returned page would include ads about tech-support and the removal of a specific strain of ransomware called ``the ICE Cyber Crime Center virus''~\cite{iceransomeware}).

%We found that a portion of the script replaces the ADs corresponding to any search query containing the word "ice" (case insensitive) to return specific technical support ADs that claim to provide services to remove the infamous~\textit{ransomeware} that locks the computer. Furthermore, it pretends to be from the "Department of Homeland Security's ICE Cyber Crime Center"~\cite{iceransomeware}. The domains and associated toll-free numbers in the ADs can be seen in the code snippet. All three are scams that we were able to verify. Full code for this can be found at INCLUDE ANONYMIZED LINK HERE.

\noindent\textbf{Domain Registration and IP:}
\label{section:registration_attribution}
The website's registration records show that it was created on 01/22/2014 to an organization called Macrofix Technical Services Private Limited which is associated to the website \texttt{macrofix.com}. This website advertises techical support services and is known to be a scam~\cite{macrofix}.

\section{Discussion and Limitations}
\label{section:discussion}

\noindent\textbf{Comparison with Past TSS studies:}
%\section{Explicit Comparison with Prior TSS Studies}
\label{section:explicit-comparison}
Given the recent work of Miramirkhani et al.~\cite{miramirkhani2017dial}, who
analyzed technical support scams and proposed a system for their discovery (ROBOVIC), 
in this section, we compare our results
with the findings of that previous study, and show that, while there is an
overlap in our findings, our TSS-discovery system allows us to find scammers
that were completely ignored by Miramirkhani et al.'s ROBOVIC.

%While our work is the first to focus on systematically understanding the TSS ecosystem 
%that originates from organic search results and sponsored advertisements on search 
%engines, a natural question to ask, is the following: does the 
%infrastructure discovered in our work have any similarities or differences with 
%those discovered by Miramirkhani et al.~\cite{miramirkhani2017dial} (M1) who use 
%domain parking and malicious ad-based url shortening services to detect TSS websites ? 
%In this section, we attempt to answer this question in two ways: first, we make a direct 
%comparison of the infrastructure found in the two studies and second we make a 
%findings-based comparison to highlight the key takeaways -- similarities and differences.

For the purpose of a direct comparison, we were able to obtain data from Miramirkhani et al. for the 
period Jan-Mar 2017, which overlaps with the second time window of data 
collection in our work. Specifically, we received a list of 2,768 FQDNs discovered by
their tool (2441 second-level domains), 882 toll-free phone numbers and 1,994 
IP addresses. Upon intersecting these sets with our own
data, we found 0/2,768 FQDNs and 0/2,441 second-level domains that were common. 
Moreover, in terms of server and telephony infrastructure, we discovered that the
two datasets had 92/1,994 common IP addresses of servers hosting TSS and 5/882 
common toll-free phone numbers. We also discovered frequent
use of ``noindex''~\cite{noindex} meta tags in the HTML source of webpages associated with 
domains in Miramirkhani et. al. dataset which was noticeably missing from 
webpages in our dataset.

Given this near-zero intersection of the two datasets, we argue that our approach
is discovering TSS infrastructure that ROBOVIC is
unable to find. Next to discovering aggressive tech support pages that ROBOVIC missed,
a core contribution of our work is focusing on ``passive'' TSS which manifest mostly as
organic search results. These pages are unlikely to be circulated over malvertising
channels: a benign-looking tech support page is unlikely to capture the attention
of users who were never searching for technical support in the first place.

Since public blacklists are still unable to capture the vast majority of TSS 
(Section~\ref{sec:blacklist-eval}), our work complements the work of 
Miramirkhani et al. Specifically, by taking advantage of
our system, blacklist curators would \emph{double} the number of TSS domains,
IP addresses, and phone
numbers that could be added to their blacklists.

\label{subsection:limitations}
\vspace{1ex}
\noindent\textbf{Limitations:} Like all real-world systems, our work is not without its limitations. Our choice
of using PhantomJS for crawling search results and ads can, in principle, be
detected by scammers who can use this knowledge to evade our monitors. We argue
that replacing PhantomJS with a real browser is a relatively straightforward task which merely
requires more hardware resources. Similarly, our choice of keeping our crawler
stateless could lead to evasions which would again be avoided if one used a
real, Selenium-driven, browser. Finally, while we provide clustering information
for the discovered TSS, in the absence of ground truth, we are unable to guarantee
the accuracy of this clustering or to attribute these clusters back to specific
threat actors.

%Inspite of the many insights and novel contributions provided in this paper around search-and-ad based 
%TSS abuse, there are clearly some limitations to our study. One of the limitations lies in data 
%collection, whereby we do not maintain the state of the scriptable headless browser designed to scrape
%the search engine listing. It is worth exploring how maintaining browser state and history (via cookies) 
%can effect the appearance of personalized TSS search and ad listings.
%In the analysis section, one limitation clearly lies in being unable to accurately attribute 
%the search-and-ad based TSS abuse/campaign to specific bad actors (people or organizations). We accept
%that this is a challenging ask and plan to explore better ways to conduct attribution in the future.

\section{Related Work}
\label{section:relwork}

As mentioned throughout this paper, Miramirkhani et al.~\cite{miramirkhani2017dial} performed the first analysis of technical support scams (TSS) by focusing on scams delivered via malvertising channels and interacting with scammers to identify their modus operandi. In recent work, Sahin et al.~\cite{relieu2017using} investigated the effectiveness of chatbots in conversing with phone scammers (thereby limiting the time that scammers have available for real users). 

The use of both the Internet and telephony channels to conduct
\textit{cross-channel} attacks is enabled by technologies that led to the convergence of
these channels. 
%Srinivasan et al.~\cite{DBLP:conf/esorics/SrinivasanGAA16} explored cross-channel abuse in 
%the context of SMS-based spam. 
Researchers have identified the evolving role of telephony and
how phone numbers play a central role in a wide range of scams, including 
Nigerian scams, phishing, phone-verified accounts, and SMS 
spam~\cite{DBLP:conf/esorics/SrinivasanGAA16,
DBLP:journals/ejisec/IsacenkovaTCFB14,DBLP:conf/esorics/SrinivasanGAA16,
DBLP:conf/ndss/GuptaSBA15,thomas2014dialing, DBLP:conf/imc/MurynetsJ12, 
Jiang:2013:GFA:2534766.2534768, Boggs:2013:DEM:2523649.2523657}.

% which is specific to technical support scams (TSS), researchers have investigated various underground ecosystems that exhibit common traits with TSS. In this section, we briefly describe the work that is the most related to ours.

%Although we focus on search-and-ad TSS abuse, other scams have used
%similar tactics and have been investigated in the past. Also, tactics specific to TSS have
%been investigated. We describe these related research efforts and discuss our contributions
%relative to them.

% Search Data Collection
%Although online pharmacy drug scams did not use the phone channel as is done
%in TSS, they used a variety of online abuse tactics, including search manipulation,
%that are used in TSS. 

In addition to telephony-specific work, researchers have analyzed a range of underground
ecosystems detailing their infrastructure and identifying the parties involved, in addition 
to potential pressure points~\cite{motoyama2010re, park2014scambaiter,
Leontiadis:2011:MAS:2028067.2028086, soska2015measuring}. Since TSS is a type of
underground ecosystem, we borrowed ideas found in prior work, such as,
the appropriate setting of \textit{User Agent} and \textit{Referrer} crawler
parameters used by
Leontiadis et al. during their analysis of drug scams~\cite{Leontiadis:2011:MAS:2028067.2028086}
to make requests appear as if they 
originated from a real user clicking on a search result. 
Also, search-redirection based
drug scams discovered by them rely on compromising high-reputation websites while the TSS
scams discovered by our system rely on black hat SEO and malicious advertisement tactics. 
Finally, there have been numerous studies that cluster abuse/spam 
infrastructure and campaigns based on URLs~\cite{DBLP:conf/sp/ThomasGMPS11}, 
IP infrastructure~\cite{DBLP:conf/uss/AntonakakisPDLF10,DBLP:conf/ndss/BilgeKKB11} 
and content~\cite{anderson2007spamscatter}. Similar hierarchical clustering techniques too have been shown effective
in multiple contexts~\cite{DBLP:conf/esorics/SrinivasanGAA16, sharma2011hicho, kapoor2014clustering, kapoor2013attribute}.
%We borrow techniques from such work to cluster TSS abuse infrastructure and do not claim novelty around the individual features or techniques used in 
%such clustering, our contribution lies in detailed analysis and investigation of of the TSS infratsructure and abuse.
% In addition, our method of %attributing the clusters by leveraging DGA footprints provides a %unique approach towards auto-labeling of SMS-spam campaigns.
In terms of countermeasures, prior work has shown the ineffectiveness of traditional blacklists 
in protecting services, such as instant messaging (IM)~\cite{DBLP:conf/ndss/PolakisPMA10}, 
and social media~\cite{DBLP:conf/ccs/GrierTPZ10, DBLP:conf/sp/ThomasGMPS11}. 
Unfortunately, until blacklist curators adopt systems such as 
our own, blacklists will also be ineffective against technical support scams.

\section{Conclusions}
\label{section:conclusion}
%Tech support scams have evolved to abuse widely used web services like search and advertising to direct victims to websites that are controlled by the scammers. Such websites use both active and passive tactics to convince unsuspecting users to call a phone number for tech support services and the phone call is used to monetize the scam. 

In this paper, we analyzed Technical Support Scams (TSS) by focusing on two new sources of scams: organic search results and ads shown next to these results. Using carefully constructed search queries and network amplification techniques, we developed a system that was able to find thousands of active TSS. We identify the presence of long-lived support domains which shield the final scam domains from search engines and shed light on the SEO tactics of scammers. In addition to aggressive scams, our system allowed us to discover thousands of passive TSS pages which appear professional, and yet display phone numbers which lead to scammers. We showed that our system discovers thousands of TSS-related domains, IP addresses, and phone numbers that are missed by prior work, and would therefore offer a marked increase of protection when incorporated into systems generating blacklists of malicious infrastructure.

\begin{sloppypar} 
\bibliographystyle{ieeetranS}
\bibliography{ms}
\end{sloppypar}
%\input{ms.bbl}
%\newpage
%{\bf Appendix}

\begin{appendix}

\begin{lstlisting}[linewidth=\columnwidth,breaklines=true,language=html, basicstyle=\tiny,frame=single,]
(int) 34 => array(
		'Goodie' => array(
			'id' => '56',
			'uid' => '1',
			'example_query' => 'ice',
			'description' => '',
			(*@ \hl{'pattern' => '/ice/i'}, @*)
			'code' => '$html = <<<EOT
<div class="search_results">
    <div class="custom-ad">
      <div class="ad_title">
        Ads related to <span class="bold">ICE Virus</span>
      </div>

      <div class="row">
        <div class="favicon"><img src=
        "https://getfavicon.appspot.com/http://www.pcvirusremove.com" height="16" width=
        "16" alt="Favicon" /></div>

        <div class="row_content">
          <div class="title">
            <a href=
            (*@ \hl{"https://goentry.com/r.php?u=http://www.pcvirusremove.com/ice-cyber-crime-center-virus-removal/">
            <b>ICE</b> Cyber <b>Virus Removal</b> - Call 1-877-635-8168 for Support.</a>} @*)
          </div>

          <div class="description">
           (*@ \hl{Expert ICE Virus Removal Help 24x7} @*)
          </div>

          <div class="link">
            <div class="link-left" style="float:left;">
              (*@ \hl{www.pcvirusremove.com} @*)
            </div>
          </div>
        </div>
      </div>

      <div class="row">
        <div class="favicon"><img src=
        "https://getfavicon.appspot.com/http://www.howtofixvirus.com" height="16" width=
        "16" alt="Favicon" /></div>

        <div class="row_content">
          <div class="title">
            <a href=
            (*@ \hl{"https://goentry.com/r.php?u=http://www.howtofixvirus.com/ice-cyber-crime-center-virus-removal/.html">
            <b>Remove ICE Virus</b> - Call US Toll Free Now!</a>} @*)
          </div>

          <div class="description">
            (*@ \hl{1-877-623-2121 ICE Virus Removal Help.} @*)
          </div>

          <div class="link">
            <div class="link-left" style="float:left;">
              (*@ \hl{www.howtofixvirus.com} @*)
            </div>
          </div>
        </div>
      </div>

      <div class="row">
        <div class="favicon"><img src=
        "https://getfavicon.appspot.com/http://ifixvirus.com" height="16" width="16" alt=
        "Favicon" /></div>

        <div class="row_content">
          <div class="title">
            <a href=
            (*@ \hl{"https://goentry.com/r.php?u=http://www.ifixvirus.com/ice-cyber-crime-virus-removal/">
            <b>ICE Virus Fix</b> | Quick Virus Removal in 2 Minutes</a>} @*)
          </div>

          <div class="description">
            (*@ \hl{Call 1-800-421-4589 for Quick fix.} @*)
          </div>

          <div class="link">
            <div class="link-left" style="float:left;">
              (*@ \hl{ifixvirus.com} @*)
            </div>
          </div>
        </div>
      </div>
    </div>
  </div><script type="text/javascript">
//<![CDATA[
  $(document).ready(function(){
  $(".sponsored_ad").hide();
  });
\end{lstlisting}

\end{appendix}

\end{document}